\documentclass[aps,prl,twocolumn,amsmath,amssymb,
longbibliography,floatfix]{revtex4-2}

\usepackage{dutchcal}
\usepackage{amsmath}
\usepackage{xcolor}
\usepackage{color}
\usepackage[normalem]{ulem}
\usepackage{float}
\usepackage{gensymb}
\usepackage{graphicx}
\usepackage{subfigure}
\usepackage{dcolumn}
\usepackage{bm}
\usepackage{setspace}
\usepackage{xcolor}
\usepackage{amsmath}
\usepackage{verbatim}
\usepackage{mathtools}
\usepackage{epstopdf}
\usepackage{lineno}
\usepackage{listings} 
\usepackage{hyperref}

\renewcommand{\topfraction}{0.9}	
\renewcommand{\bottomfraction}{0.8}	
\renewcommand{\dbltopfraction}{0.9}	
\renewcommand{\textfraction}{0.07}	
\renewcommand{\floatpagefraction}{0.7}	
\renewcommand{\dblfloatpagefraction}{0.7}	

\graphicspath{ {images/} }

\begin{document}

\title{
Effective particle size governs structure and dynamics in rough hard-particle fluids
}

\author{Nanqing Xiao$^1$}
\author{Zhen Zhang$^1$}
\author{Walter Kob$^{1,2}$}
\email[Corresponding author: ]{walter.kob@umontpellier.fr}
\author{Yujie Wang$^{1,3,4}$}
\email[Corresponding author: ]
{yujiewang@sjtu.edu.cn}
\affiliation{$^1$ College of Physics, Chengdu University of Technology, Chengdu 610059, China}
\affiliation{$^2$ Department of Physics, University of Montpellier, CNRS, F-34095 Montpellier, France}
\affiliation{$^3$ State Key Laboratory of Geohazard Prevention and Geoenvironment Protection, Chengdu University of Technology, Chengdu 610059, China}
\affiliation{$^4$ School of Physics and Astronomy, Shanghai Jiao Tong University, Shanghai 200240, China}
\date{\today}

\begin{abstract}

We numerically investigate how particle surface roughness affects the static and dynamic properties of a hard-particle fluid across a wide range of densities, $\rho$. These simulations of a simple model of granular systems reveal that, although the amplitude and coverage of surface corrugation significantly influence the $\rho$-dependence of system properties, the qualitative behavior of this dependence remains unchanged. These findings can be described quantitatively by introducing an effective particle size, which enables a direct mapping of the rough particle systems to equivalent hard-disk systems. Analytical calculations provide an explicit form of this mapping, and allow us to predict the static and dynamic properties of rough particles using standard liquid-state theories for hard disks. 
\end{abstract}

\maketitle 
\clearpage

Granular systems are ubiquitous in nature and technological applications since they are closely related to phenomena like landslides, avalanches, or food and powder processing~\cite{deGennes1999,Jaeger1996,ForterrePouliquen2008,Iverson1997,Saleh2018}. 
These materials are inherently athermal, frictional, and dissipative, resulting that they can have highly complex structures and relaxation dynamics, making them paradigmatic examples that allow to explore nonequilibrium physics~\cite{GDRMiDi2004,Duran2012,Andreotti2013,Franklin2016,Anthony2005JGR}. 

Recent studies have advanced our understanding regarding the influence of friction on the static properties of the packs, notably the phenomenon of jamming and the details of the local packing geometry~\cite{Shundyak2007,Tang2023,Tang2025,Silbert2010,Santos2020,Somfai2007}.  
However, its effect on dynamic quantities is still an open problem since the induced coupling between the translational degrees of freedom and the orientational ones renders the particle motion highly complex~\cite{Kou2018,Brilliantov2007,Megias2023,Trittel2024}. 
Also surface roughness is an important feature of granular systems since it has been shown to affect the static and dynamical properties of the system in a similar manner as friction~\cite{Yuan2024NatComm,Hsu2018,Ilhan2022,Kato2023,Hu2020Langmuir,Hsiao2019COCIS}. However, their origin is fundamentally different: Friction is due to the tangential resistive force at contacts, influencing sliding and rolling \cite{Papanikolaou2013,Silbert2010}, whereas geometric roughness introduces steric constraints and modifies the excluded volume even in the absence of tangential forces \cite{Franklin2012Geometric}. So far, insight on the connection between surface roughness and friction is still missing since theory struggles to describe the properties of rough particles, while in experiments it is difficult to vary the roughness in a systematic manner, although recent advances in 3D printing techniques have allowed to address this problem~\cite{Ikeda2020,Lu2025ColCom,Kronenfeld2024Nature,Xu2021ComposB}.

Simple models that allow to tune the roughness show that the jamming density and coordination number decrease with roughness and converges to the behavior of the frictional case~\cite{Ikeda2020}, but the important case of densities beyond this point has so far remained unexplored. Previous computational studies examined the jamming of various types of frictionless nonspherical smooth particles and showed that the packs can remain mechanically stable even if the average contact number is inferior than the number of degrees of freedom, thus demonstrating that a non-trivial particle shape can strongly influence the properties of the pack~\cite{VanderWerf2018,Marschall2015Compression,Aponte2024Geometric}. Rheological experiments demonstrate that rough particles exhibit more pronounced discontinuous shear thickening and an earlier onset of dilatancy compared to their smooth counterparts, emphasizing the importance of asperity-induced solid–solid contacts~\cite{Hsiao2019Review,Ikeda2020,Hu2020Langmuir, Hsiao2019COCIS,Murphy2019GranulMatter,Goldhirsch2005PRL,Gayen2008PRL,Lootens2005PRL,Hsiao2017PRL,Hsu2021NatComm}. 
While these findings demonstrate the importance of roughness for the macroscopic behavior of the system, we lack at present a theoretical framework that allows to describe such systems in a wider range of densities, which also hinders to establish the connection between friction and surface roughness in these systems. 

Here we present the results of large scale computer simulations of systems of rough particles. We find that in a wide range of densities, their structural and dynamic properties can be mapped onto a hard-disk fluid with an effective particle size that can be predicted analytically, thus allowing well known liquid state theory approaches to describe these complex systems.

\begin{figure}[tbp]
\centering
\subfigure{
\includegraphics[width=0.48\textwidth]{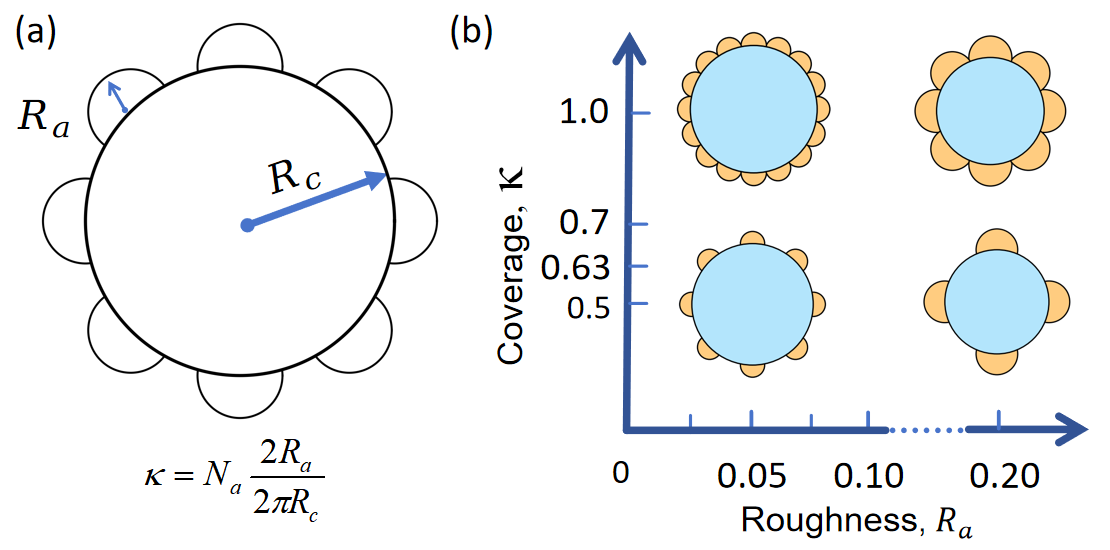}
}
\caption{
(a) Schematic representation of a rough particle with a central core of radius $R_c$ and surface asperities of radius $R_a$ uniformly distributed along the perimeter. Surface coverage is defined as $\kappa=N_a R_a/R_c$, where $N_a$ is the number of asperities. 
(b) Examples of particles with different roughness $R_a$ and coverage $\kappa$.
}
\label{fig_par}
\end{figure}

To probe how the properties of a fluid depend on particle roughness, we consider a two dimensional model of disks that are decorated by asperities~\cite{Papanikolaou2013}. Using the diameter, 2$R_c$, of the central particle as unit length, the relevant parameters characterizing a meso-particle (MP) are the radius of the asperities, $R_a$, and the fraction of the perimeter of the central particle that is covered by the latter, $\kappa$, see Fig.~\ref{fig_par}. We consider $0 \leq R_a \leq0.2$ and for $\kappa$ the values 1.0, 0.7, 0.63, and 0.5. 
MPs interact with each other by means of a Hertzian central force~\cite{Papanikolaou2013}, i.e., $F_{\rm HZ}=[R_i R_j/(R_i+R_j)]^{1/2} k_n \delta^{3/2}$, where $\delta$ is the overlap between two disks of radius $R_i \in \{ R_c, R_a\}$ and $k_n=2 \times 10^8$, a value that is sufficiently large to ensure that the MP are very similar to hard bodies. The total number of MP was $10^4$, large enough to avoid significant finite size effects but small enough that Mermin-Wagner fluctuations do not significantly affect the system properties on the length and time scales considered here~\cite{BinderKob2011}. More details on the simulations are given in the Appendix.

The structural order of a fluid is usually characterized by means of the radial distribution function $g(r)$~\cite{HansenMcDonald2013}, see SM Fig.~S1.
However, the nature of the local order can be studied better by means of a three-point correlation function, such as the bond-orientational order $\overline{\psi}_6$~\cite{NelsonHalperin1979}, which is the system average of the local sixfold bond-orientational order parameter $\psi_6(k)$, defined for particle $k$ as

\begin{equation}
\psi_6(k)=\left|\frac{1}{N_{k}}\sum_{j=1}^{N_{k}} \exp\!\left[\,i 6\theta_{kj}\right]\right|,
\end{equation}

\noindent
where the sum runs over the $N_{k}$ Voronoi neighbors of particle $k$, and $\theta_{kj}$ is the angle between the bond vector $\mathbf{r}_{kj}$ and a fixed reference axis. 
\begin{figure}[b]
    \centering
    \includegraphics[width=0.48\textwidth]{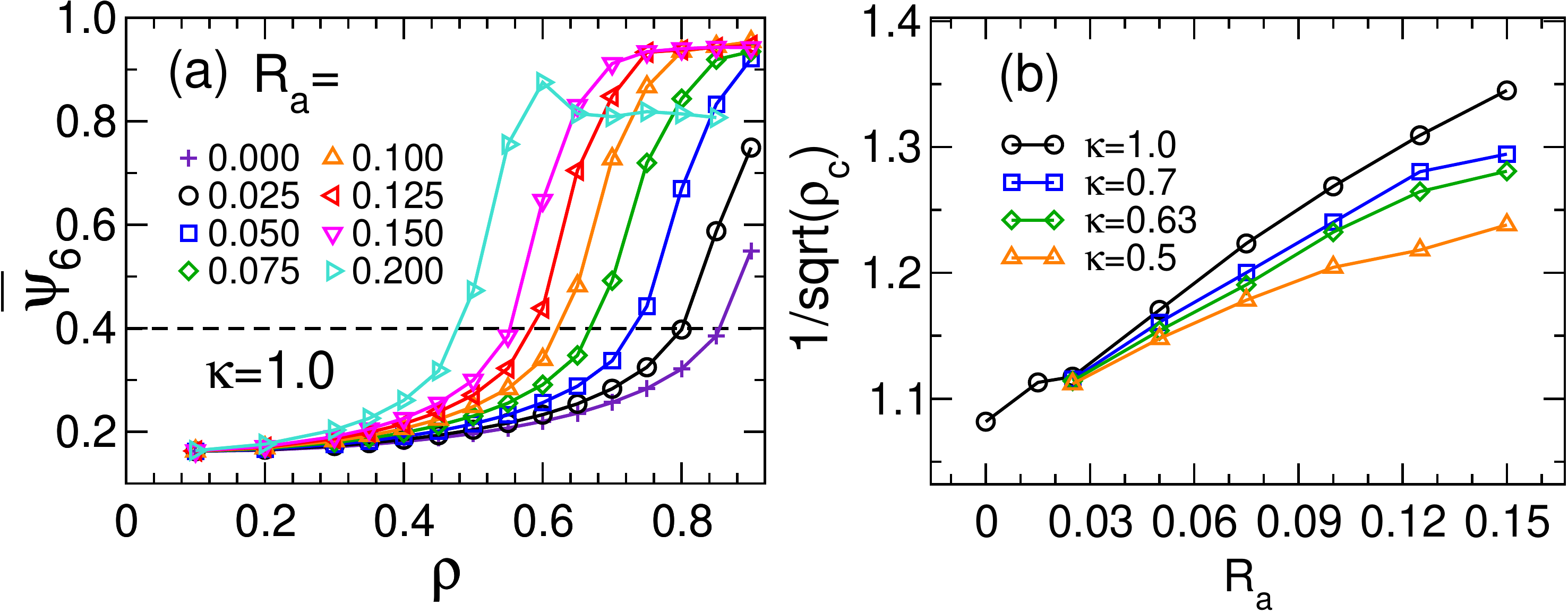}
    \caption{(a) Bond-orientational order parameter $\overline{\psi}_6$ as a function of number density $\rho$ for $\kappa=1.0$ and different asperity radii $R_a$. The horizontal dashed line shows $\overline{\psi}_6(\rho_c)=0.4$ and serves to define $\rho_c$ presented in panel (b).
    (b) $1/\sqrt{\rho_c}$ as a function of $R_a$ for different values of $\kappa$.}
    \label{fig_q6_A}
\end{figure}

\begin{figure}[th]
\centering
\includegraphics[width=0.48\textwidth]{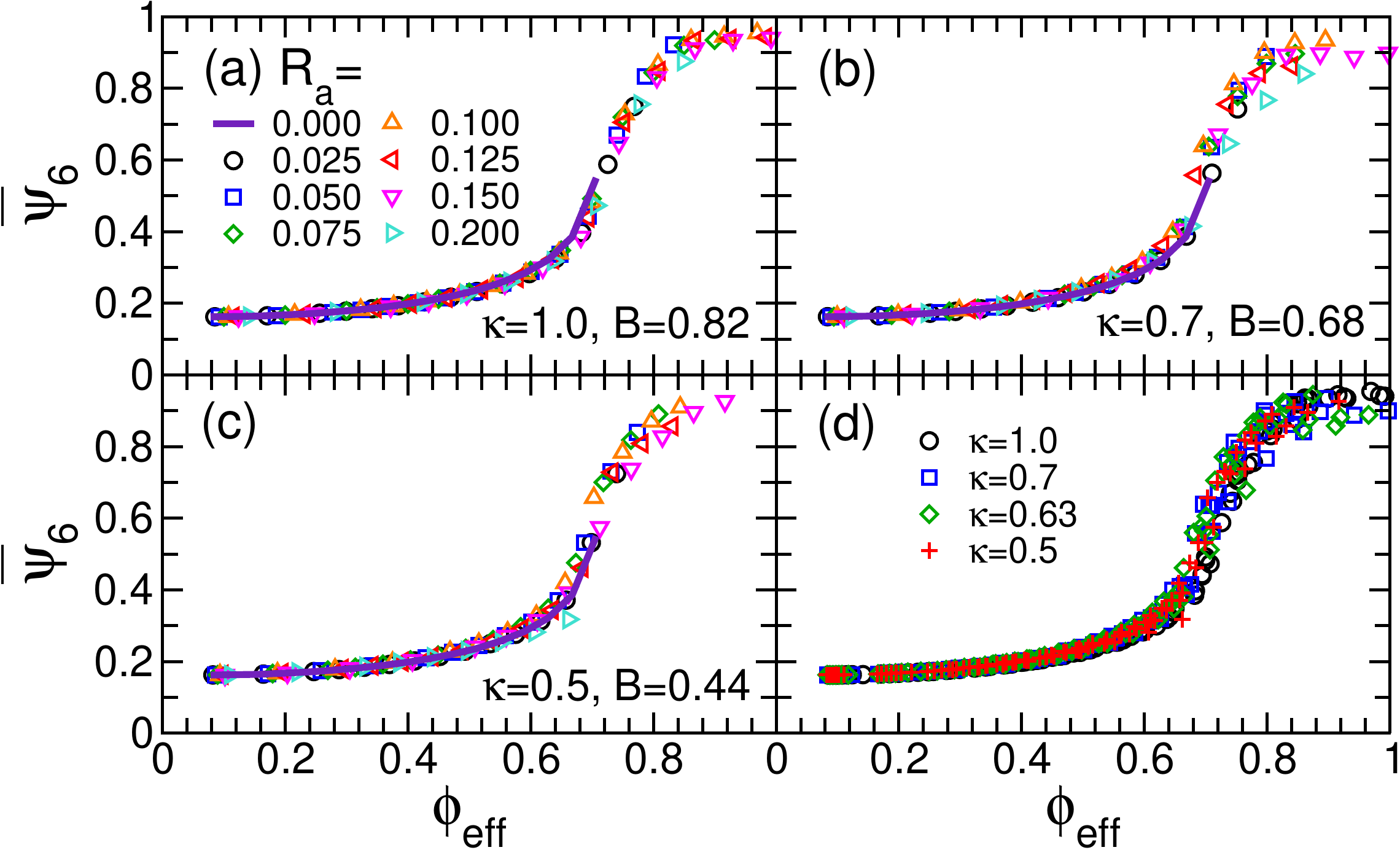}
\caption{Bond-orientational order parameter $\overline{\psi}_6$ as a function of the effective packing fraction $\phi_{\rm eff}$ for different values of the asperity radius $R_a$. Panels (a)--(c) correspond to $\kappa=1.0$, $0.7$, and
$0.5$, respectively, while panel (d) combines the data for all $\kappa$
($\kappa=1.0, 0.7, 0.63$, and $0.5$) to illustrate the resulting master-curve collapse. For each $\kappa$, the factor $B$ used to make the master curve is indicated in the panel. Note that we show only data points for which the samples have not crystallized. 
}
\label{fig_q6_phi}
\end{figure}

Figure~\ref{fig_q6_A}(a) shows the $\rho$-dependence of $\overline{\psi}_6$ (see Figs.~S2 and S3 for the PDF of $\psi_6$). 
At low densities $\overline{\psi}_6$ is small, but increases quickly at intermediate densities, reflecting the formation of a local order, before it saturates at high $\rho$ due to the formation of intermediate range order in the form of poly-crystals. 
The shape of the curves for the different $R_a$ is independent of $R_a$, and only the density at which $\bar{\psi}_6$ shoots up depends on $R_a$. This suggests that systems with different $R_a$ can be mapped onto each other and this conclusion holds also for other values of coverage $\kappa$ or local structural observables (see SM).

For this mapping we introduce an effective radius $R_{\rm eff}$\\[-10mm]

\begin{equation}
R_{\rm eff}(R_a,\kappa) = R_c(1+B(\kappa) R_a/R_c) \quad .
\label{eq1}
\end{equation}

\noindent
Here $B(\kappa)$ is a fit-parameter that depends on $\kappa$ and determines the effective size of a meso-particle. The effective packing fraction is then given by

\begin{equation}
    \phi_{\rm eff} = \rho \pi R^2_{\rm eff} \quad.
    \label{eq_phieff}
\end{equation}

\noindent
To obtain $B$, we determine for a given $R_a$ the density $\rho_c$ at which $\overline{\psi}_6$ takes the value 0.4, horizontal dashed line in Fig.~\ref{fig_q6_A}(a). (The exact value of this threshold is not important.) Assuming that $\rho_c$ corresponds to a fixed $\phi_{\rm eff}$, one obtains from Eq.~(\ref{eq_phieff}) that $1/\sqrt{\rho_c} \propto R_{\rm eff}$. 
Figure~\ref{fig_q6_A}(b) shows that $1/\sqrt{\rho_{c}}$ is indeed a linear function of $R_a$ if $R_a/R_c$ is small, i.e., weak roughness, with the slope of the curves giving $B(\kappa)$.

Figure~\ref{fig_q6_phi}(a)-(c) presents $\overline{\psi}_6$ as a function of $\phi_{\rm eff}$ and one sees that the different data sets fall perfectly on a master-curve, demonstrating that one can indeed map the properties of systems with different $R_a$ onto each other by means of Eq.~(\ref{eq1}).
The figure hints that the master curves are basically independent of the coverage $\kappa$. That this is indeed the case is demonstrated in Fig.~\ref{fig_q6_phi}(d) where we superimpose the data for the different values of $\kappa$. From this plot one thus concludes that the $\rho$-, $R_a$-, and $\kappa$-dependence of the local structure, as characterized by $\overline{\psi}_6$, can be obtained from a single function that depends only on $\phi_{\rm eff}$. 
In the SM we show that this conclusion holds not only for $\overline{\psi}_6$ but for other observables as well, such as the bond-angle distribution or the local order parameter $T_6$~\cite{Tong2018HiddenOrder}.

\begin{figure}[ht]
    \centering
    \includegraphics[width=8cm]{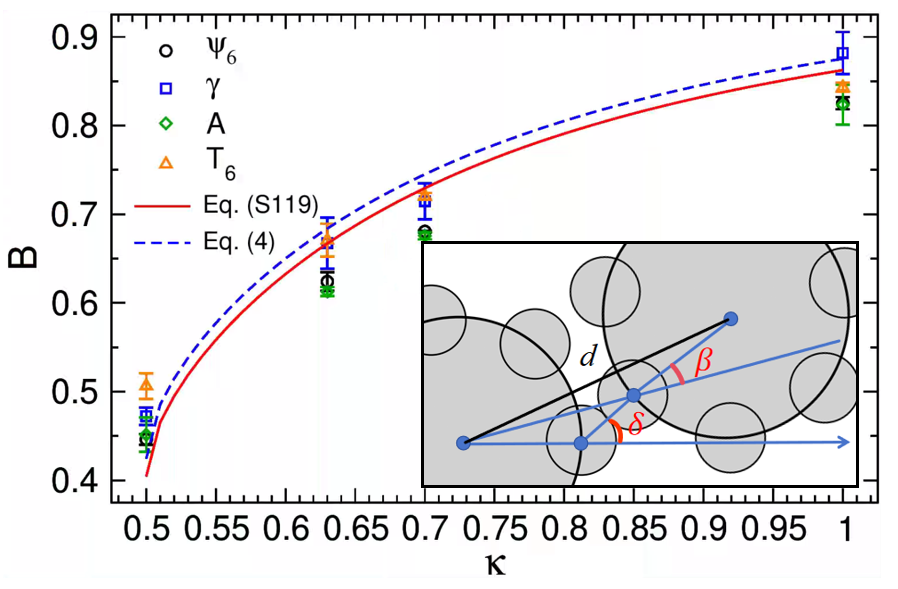}
    \caption{Scaling factor $B$ as a function of the surface coverage ratio $\kappa$, extracted from four different structural quantities: The bond-orientational order parameter $\overline{\psi}_6$, the Lindemann ratio $\gamma$, the bond-angle distribution $A$, and the triplet orientational order parameter $T_6$ (see SM for definitions). The red solid line shows the result of our analytical calculation (Eq.~S119 in the SM), while the blue dashed line corresponds to its first-order approximation, Eq.~(\ref{eq_B_approx}). Inset: Illustration of the generic contact geometry between two MPs: The relative rotation of the right particle around the contact point (angles $\beta$ and $\delta$) yields all configurations needed to calculate the effective distance $\bar{d}$ between the two centers, see the SM for details. 
}
    \label{fig_B_kappa}
\end{figure}

The mapping of the data for a given observable and fixed $\kappa$ onto a common master curve might a priori depend on the observable considered. That this is not the case is shown in Fig.~\ref{fig_B_kappa} which shows that the $\kappa$-dependence of $B$ for the different observables is basically independent of the quantity considered. In particular we present the value of $B$ as obtained from making scaling plots like the ones shown in Fig.~\ref{fig_q6_phi} for the observables $\overline{\psi}_6$, from the area $A$ under the main peak in the bond-angle distribution function $\theta_{ijk}$ between three particles that are mutual nearest neighbors, the local orientational order parameter $T_6$, and the size of the cage formed by the nearest neighbors or a particle, i.e., the Lindemann ratio (see SM for a precise definition of these quantities). That $B(\kappa)$ increases monotonically with the coverage, is consistent with the expectation that particles with larger $\kappa$ possess a more complete ring of asperities and thus a stronger excluded–volume effect, i.e., a larger zone in which the repulsive interaction is very large. The independence of $B$ from the observable considered leads to the conclusion that the mapping given by Eq.~(\ref{eq1}) allows to obtain a universal, i.e., observable-independent, master curve, with $B$ being the only unknown ($\kappa$-dependent) parameter. In other terms, once one has obtained this master curve, for example for the case of smooth hard disks, one can predict via Eq.~\ref{eq1} the properties of all the systems with $R_a>0$ and $\kappa>0$.

\begin{figure}[ht]
    \centering
    \includegraphics[width=0.48\textwidth]{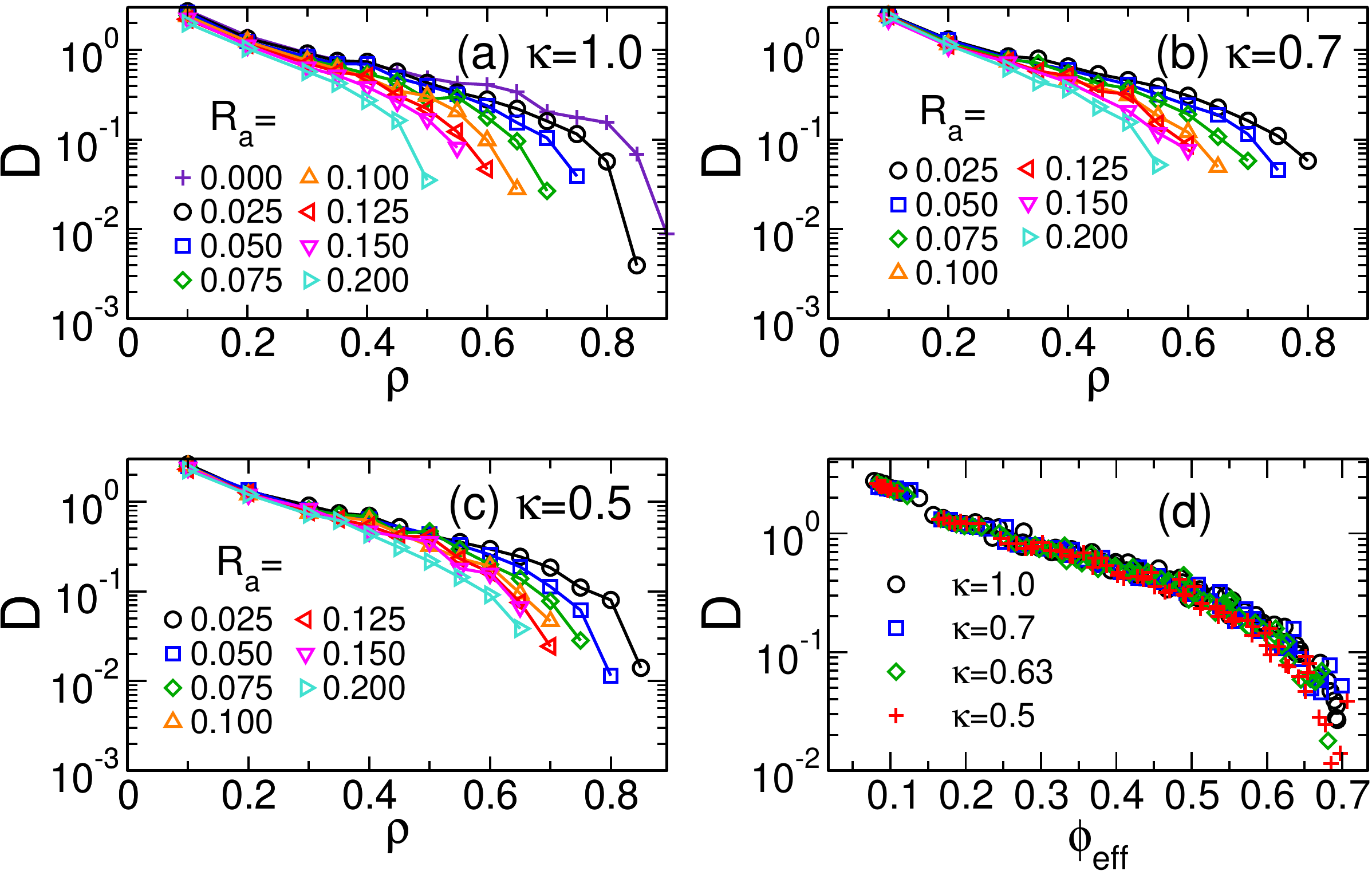}
    \caption{Diffusion constant $D$ as a function of number density $\rho$ for different asperity radii $R_a$ at fixed $\kappa$: (a) $\kappa=1.0$, (b) $\kappa=0.7$, and (c) $\kappa=0.5$. (d) The same data plotted versus the effective packing fraction $\phi_{\mathrm{eff}}$, showing a collapse onto a master curve.
    }
    \label{fig_MSD_crMSD}
\end{figure}

The independence of $B$ from the observable considered hints that the static properties of the systems can be obtained from an analytical calculation that takes into account the geometry of the particles. Since this is basically a hard core system, all configurations have the same Boltzmann weight and thus one is faced only with a counting problem to obtain the entropic contribution. For the case of a hard-disk system this is a formidable problem that is still open, but many excellent approximations exist~\cite{ParisiZamponi2010RMP,Royall2024}. We now assume that the partition function of our system can be factorized into the one of a hard disk system and a second factor that depends only on the orientational degrees of freedom. To evaluate this second factor we consider two MPs that touch each other and calculate their average distance $\bar{d}$ as a function of $R_a$ and $\kappa$. Assuming that $\bar{d}$ is a reliable proxy for $R_{\rm eff}$, this calculation allows us to obtain $B$ from Eq.~(\ref{eq1}). 

For the calculation of $\bar{d}$, one has to take the average over all distances $d$ between the centers of the two touching MPs, which in turn depend on the relative orientations of the MPs, i.e., the angles $\beta$ and $\delta$ defined in the inset of Fig.~\ref{fig_B_kappa}.
These calculations are presented in the SM and differentiating the resulting $\bar{d}$ with respect to $R_a$ gives $B$, Eq.~(S119). 
By making the approximation that $R_a/R_c \ll 1$ one obtains the expression, see Eq.~(S120),
\begin{equation}
    B \approx \frac{1}{4\kappa\arcsin(\frac{1}{2\kappa})}\left(2-\frac{1}{6\kappa^2}\right)
    \label{eq_B_approx} \quad,
\end{equation}
and both results are included in Fig.~\ref{fig_B_kappa} as well. The good agreement between theory and data from the simulation shows that it is indeed possible to predict the static properties of the systems of MPs if one knows the ones of a hard-disk system.

In disordered systems that relax slowly, the dynamics is closely related to the structure~\cite{BinderKob2011}, and hence it is of interest to probe whether the effective size $R_{\rm eff}$ of the MPs allows to predict also the relaxation dynamics of the system. To check this we have determined the mean-squared displacement (MSD) of a tagged particle and, using the Einstein relation, calculated the diffusion constant $D$. (We recall that in 2D systems the long time diffusion constant is not well defined because of the long-time tails in the velocity auto-correlation function. However, as it is custom in the field of glass-physics, we consider here only MSDs that are much smaller than the size of the system squared, i.e., $D$ is not affected in a noticeable manner by these tails.) Figure~\ref{fig_MSD_crMSD}(a)-(c) presents the density dependence of $D$ for different values of $R_a$ and $\kappa$. As expected, and independent of $\kappa$, the diffusion coefficient decreases rapidly with increasing density, reflecting the strong dynamical slowing down caused by the buildup of local constraints. We note that also here the overall shape of the curves is independent of $R_a$, although the density at which the slowdown becomes pronounced changes with $\kappa$ and $R_a$. Particles with larger coverage ratios or larger asperities exhibit a reduction in mobility already at lower densities, consistent with the enhanced excluded-volume effects discussed in the structural analysis.

To examine whether also the dynamical properties are governed by the effective size $R_{\rm eff}$, we plot in Fig.~\ref{fig_MSD_crMSD}(d) the diffusion constant $D$ as a function of the effective packing fraction $\phi_{\mathrm{eff}}$. Remarkably, the curves for the different $R_a$ and $\kappa$ collapse onto a single master curve, demonstrating that $R_{\rm eff}$ controls also the translational dynamics. 
The presented results show that the structural and dynamical properties of two-dimensional particles with different surface roughness can be mapped onto a hard-disk system with an effective size $R_{\rm eff}$ and hence an effective packing fraction $\phi_{\rm eff}$. This mapping works even for a roughness that is very substantial, i.e., 20\%. Our analytical calculation allows to predict with good accuracy this mapping as a function of $R_a$ and $\kappa$ thus opening the door to describe the static and dynamic properties of such systems from the knowledge of the well-studied model of hard disks. Hence we conclude that for a large range of density, roughness, and asperity coverage (i) geometric roughness primarily renormalizes the excluded volume and thus controls a broad class of structural and transport observables via a single $R_{\rm eff}$, and (ii) the coverage dependence $B(\kappa)$ can be computed from the geometrical properties of the MPs, enabling quantitative predictions from the hard-disk reference system.
This insight will allow to make a better connection between the concepts of friction and roughness, a relation that is presently not well understood~\cite{Hsiao2019Review,Ikeda2020}.

Whether the proposed approach can be successfully generalized to 3D systems is at present an open question, but the mean-field nature of the calculation leads to the expectation that this can indeed be done. To what extent this approach is also reliable in the case of sparse asperities coverage, or if these lockings are strongly orientational, or in the case of multi-component systems, remains to be tested in the future. 
A further important question is whether this approach is applicable if the overall shape of the particle is non-spherical (elliptic, kidney-shaped,...), i.e., 
particle geometries that are often found in real granular materials~\cite{Jaeger1996}. At high densities the free energy landscape of such systems can be expected to become significantly more complex than the one of smooth circular objects, with the presence of roughness giving rise to new relaxation channels~\cite{Yuan2024NatComm,Hsiao2019COCIS,Murphy2019GranulMatter}. These novel relaxation mechanisms will in turn affect the static and dynamic properties of the system in the jammed and glassy state, i.e., regimes that are so far not well understood. Hence it will be of interest to probe to what extent the present mean-field like calculation is able to capture these behaviors.

Acknowledgments: We thank H. Yuan for a critical reading of the manuscript. This work was supported by the National Natural Science Foundation of China (Grant No. 12534008) and the Space Application System of China Manned Space Program (Grant No. KJZ-YY-NLT0504).\\[2mm]

Author contributions: Z.Z. and W.K. conceptualized the work; N.X. carried out the simulations and did the analysis of the data; all authors interpreted the results and wrote the paper.

\section{Reference}
\bibliography{references}

\vspace*{20mm}
{\bf Appendix}\\[1mm]

{\it Model:} We consider a one-component granular system composed of meso-particles (MPs), each consisting of a central disk of radius $R_c$ and $N_a$ small disks of radius $R_a$ that have their center  on the surface of the central disk. The diameter of the central disk will be used as the unit of length. The surface coverage $\kappa$ quantifies what fraction of the perimeter of the central disk is occupied by these asperities, with $\kappa = 1.0$ corresponding to complete coverage. In the following we will consider $\kappa = 0.5$, $0.63$ (maximal coverage that allows an asperity to contact the central disk), $0.7$, and $1.0$. The radius $R_a$ of the asperities characterizes the roughness of the meso-particle~\cite{Papanikolaou2013}. To explore different roughness levels, we vary $R_a$ in the range $0.025$–$0.2$.

The number of asperities is given by

\begin{equation}
N_a = \left\lfloor \frac{2\pi R_c}{2R_a}\kappa \right\rfloor ,
\label{eq_Ne}
\end{equation}

\noindent
where $\lfloor \cdot \rfloor$ denotes the integer floor function. This value of $N_a$ can result in a coverage $\kappa$ that is slightly smaller than the target value and in that case we have slightly decreased $R_a$ to reach a given $\kappa$.
These adjustements are typically less than 2\% for $\kappa=1.0$, but can be larger for certain combinations such as $\kappa=0.5$ and $R_a=0.15$, see Table~\ref{table1}. 

\begin{table}[h]
    \centering
    \resizebox{7cm}{!}{
    \begin{tabular}{|c|c|c|c|c|}
        \hline
        & \(\kappa = 0.5\) & \(\kappa = 0.63\) & \(\kappa = 0.7\) & \(\kappa = 1.0\) \\
        \hline
        \(R_a = 0.025\) & 0.0245 & 0.0247 & 0.0250 & 0.0250 \\
        \hline
        \(R_a = 0.050\) & 0.0491 & 0.0495 & 0.0500 & 0.0499 \\
        \hline
        \(R_a = 0.075\) & 0.0714 & 0.0707 & 0.0733 & 0.0748 \\
        \hline
        \(R_a = 0.100\) & 0.0982 & 0.0990 & 0.1000 & 0.0982 \\
        \hline
        \(R_a = 0.125\) & 0.1122 & 0.1237 & 0.1221 & 0.1221 \\
        \hline
        \(R_a = 0.150\) & 0.1309 & 0.1414 & 0.1374 & 0.1428 \\
        \hline
        \(R_a = 0.200\) & 0.1964 & 0.1980 & 0.1833 & 0.1964 \\
        \hline
    \end{tabular}
    }
    \caption{The exact value of $R_a$ used in the simulations to assure a constant value of coverage $\kappa$.}
    \label{table1}
\end{table}

\begin{table}[h]
    \centering
    \resizebox{7cm}{!}{
    \begin{tabular}{|c|c|c|c|c|}
        \hline
        & \(\kappa = 0.5\) & \(\kappa = 0.63\) & \(\kappa = 0.7\) & \(\kappa = 1.0\) \\
        \hline
        \(R_a = 0.025\) & 32 & 40 & 44 & 63 \\
        \hline
        \(R_a = 0.050\) & 16 & 20 & 22 & 31 \\
        \hline
        \(R_a = 0.075\) & 11 & 14 & 15 & 21 \\
        \hline
        \(R_a = 0.100\) & 8 & 10 & 11 & 16 \\
        \hline
        \(R_a = 0.125\) & 7 & 8 & 10 & 13 \\
        \hline
        \(R_a = 0.150\) & 6 & 7 & 8 & 11 \\
        \hline
        \(R_a = 0.200\) & 4 & 5 & 6 & 8\\
        \hline
    \end{tabular}
    }
    \caption{The number of asperities for the different simulated systems.}
    \label{The number of the bump table 2}
\end{table}

\begin{table}[h]
    \centering
    \resizebox{7cm}{!}{
    \begin{tabular}{|c|c|c|c|c|}
        \hline
        & \(\kappa = 0.5\) & \(\kappa = 0.63\) & \(\kappa = 0.7\) & \(\kappa = 1.0\) \\
        \hline
        \(R_a = 0.025\) & 0.8153 & 0.8237 & 0.8286 & 0.8475 \\
        \hline
        \(R_a = 0.050\) & 0.8465 & 0.8633 & 0.8730 & 0.9088 \\
        \hline
        \(R_a = 0.075\) & 0.8754 & 0.8979 & 0.9154 & 0.9755 \\
        \hline
        \(R_a = 0.100\) & 0.9107 & 0.9450 & 0.9647 & 1.0373 \\
        \hline
        \(R_a = 0.125\) & 0.9296 & 0.9870 & 1.0066 & 1.0983 \\
        \hline
        \(R_a = 0.150\) & 0.9550 & 1.0175 & 1.0359 & 1.1585 \\
        \hline
        \(R_a = 0.200\) & 1.0468 & 1.1180 & 1.1257 & 1.3094 \\
        \hline
    \end{tabular}
    }
    \caption{The surface area of the complete MP.\\
    }
    \label{area of MP table 3}
    \end{table}

In order to compare our results with hard disk-like results, we also simulate a smooth-particle model with $R_a = 0$. To maintain identical simulation protocols (e.g., initialization and temperature control), these particles are implemented by attaching two tiny asperities with $R_a = 10^{-4}$ on opposite sides of the central disk. This minimal perturbation ensures negligible impact on the particle’s geometry and dynamics. In all cases, the total mass of each MP is normalized to unity by adjusting the mass density. (The surface of a MP is given in Table III.) Denoting by $N$ the number of meso-particles, the number of interaction sites used in the simulation is given by $N(N_a+1)$, which ranges between 50,000 and 640,000 (see Table II).\\[2mm]

{\it Interaction potential:}
The particle interactions in our simulations are given by the Hertzian interaction model, i.e., the interaction force, $\mathbf{F}_{\rm HZ}$, is given by \cite{Mindlin1949,Silbert2001}

\begin{align}
\mathbf{F}_{\rm HZ} = \sqrt{\frac{R_iR_j}{R_i+R_j}} k_n \delta^{1.5} \mathbf{n}_{ij} \quad .
\label{eq_interaction}
\end{align}

\noindent

Here $\delta$ denotes the overlap between two disks $i$ and $j$, i.e., $R_i \in \{R_c, R_a\}$, and $\mathbf{n}_{ij}$ is the unit vector connecting their centers.

We present the results in dimensionless units, with $\sigma=2R_c$ the unit of length, the mass of a MP as $m=1.0$, and time in units of $\tau = \sqrt{\frac{m}{k_n\sigma}}$. The parameter $k_n$ is chosen as $k_n = 2 \times 10^8$, a value that ensures that the particles are very similar to frictionless hard core bodies, i.e.,  the dynamics is strongly dominated by excluded-volume effects and the presence of surface asperities. A similar choice of the parameters has been made in previous studies of sheared granular systems, where they were shown to reproduce realistic relaxation processes while keeping particle overlaps negligible~\cite{Mao2022,Mao2024}.\\[2mm]

{\it Simulation Details:}
Simulations are performed using the Large-scale Atomic/Molecular Massively Parallel Simulator (LAMMPS)~\cite{Plimpton1995LAMMPS,Thompson2022LAMMPS}. 
Periodic boundary conditions are applied in all three spatial directions, with the $z$-dimension fixed to $0.5$ to ensure quasi-two-dimensional confinement. 
The initial configurations are prepared in a $320 \times 320 \times 0.5$ simulation box by randomly placing $N=10\,000$ MPs, each one having been assigned a random velocity corresponding to the target temperature of $T=1.0$. (Since the interaction potential is very stiff, we basically simulate hard disks and hence the value of the temperature is not a relevant quantity.) The box dimensions in the $x$–$y$ plane are then slowly reduced to achieve the desired number density $\rho$.

The system is equilibrated in the canonical ($NVT$) ensemble using a Nosé–Hoover thermostat \cite{Nose1984,Hoover1985} with a target temperature $T=1.0$ and a thermostat damping time of $\tau_{\mathrm{damp}} = 0.5$. Each simulation consisted of an equilibration stage of $3\times 10^{6}$ steps, followed by a production run of $7\times 10^{6}$ steps for data collection. 
The time step was $\Delta t = 1.0 \times 10^{-4}$, a value that is small because the potential is very stiff. We note that the equilibration time is chosen to be at least one order of magnitude larger than the structural relaxation time at the highest simulated density, thereby ensuring that the system reaches equilibrium before measurements are taken.
At the highest densities the relaxation dynamics of the system becomes very slow due to the formation of a polycrystalline structure. In order to avoid out-of-equilibrium effects we present only data that that is not significantly influenced by these effects. 
\\

\newpage
\begin{widetext}

\renewcommand{\topfraction}{0.9}	
\renewcommand{\bottomfraction}{0.8}	
\setcounter{topnumber}{2}
\setcounter{bottomnumber}{2}
\setcounter{totalnumber}{4}     
\setcounter{dbltopnumber}{2}    
\renewcommand{\dbltopfraction}{0.9}	
\renewcommand{\textfraction}{0.07}	
\renewcommand{\floatpagefraction}{0.7}	
\renewcommand{\dblfloatpagefraction}{0.7}	
	
\renewcommand{\figurename}{Fig.}
\renewcommand{\thefigure}{S\arabic{figure}}
\setcounter{figure}{0}

\renewcommand{\theequation}{S\arabic{equation}}
\setcounter{equation}{0}

\section{Supplementary Material}

\noindent
{\bf Effective particle size governs structure and dynamics in rough hard-particle fluids}

Nanqing Xiao, Zhen Zhang, Walter Kob, and Yujie Wang\\[2mm]

\subsection{Radial distribution function and static structure factor}

\begin{figure}[ht]
\centering
\includegraphics[width=\linewidth]{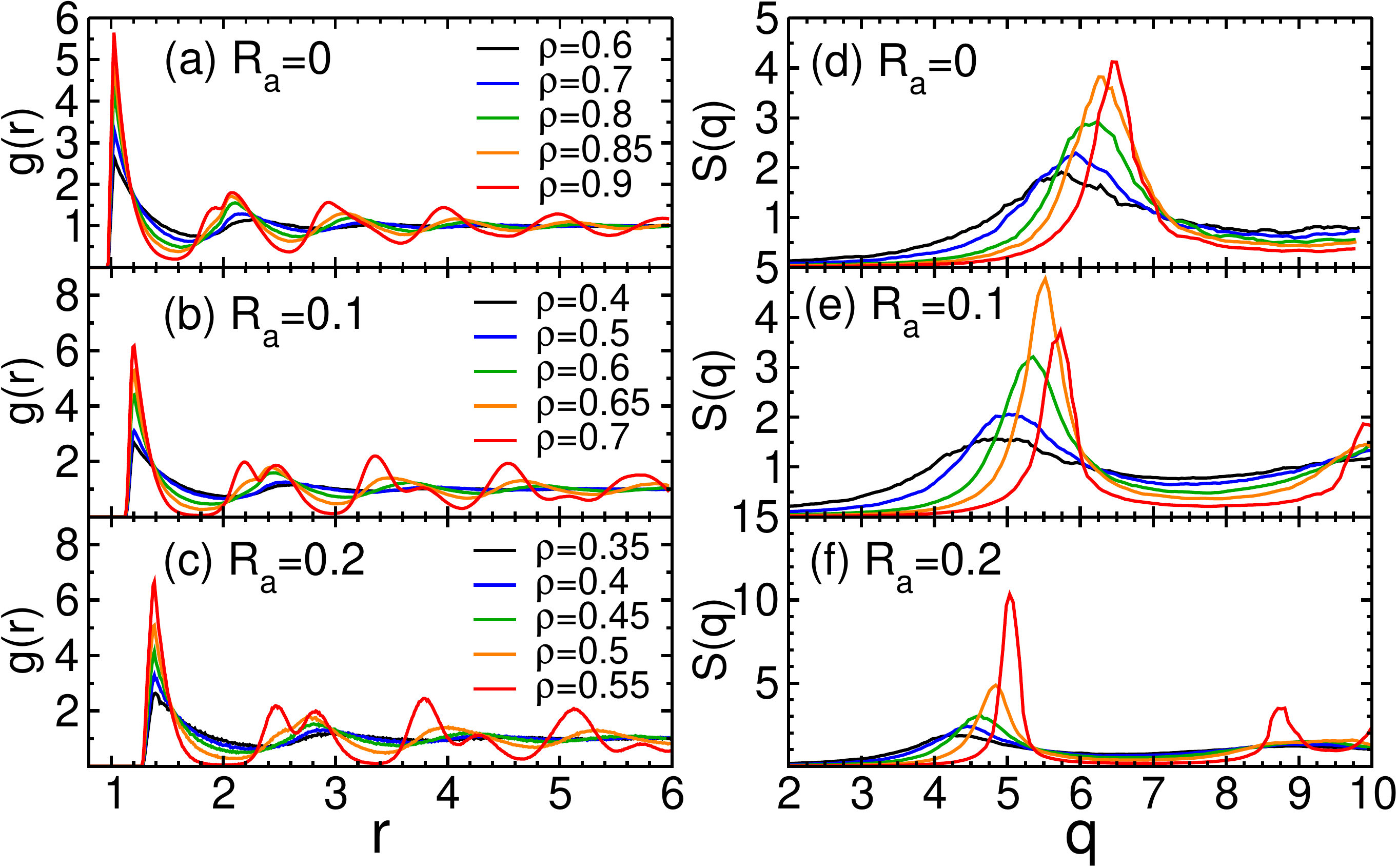}
\caption{Radial distribution function $g(r)$ (left) and static structure factor $S(q)$ (right) for the one-component two-dimensional rough-particle system with $\kappa=1.0$ at different roughnesses $R_a$ and number densities $\rho$. From top to bottom: $R_a=0$, $0.1$, and $0.2$. With increasing $\rho$, the first peak of $g(r)$ and the main peak of $S(q)$ become more pronounced, indicating enhanced translational correlations. 
}
\label{fig_gr_sq}
\end{figure}

The pair distribution function $g(\mathbf{r})$ and static structure factor $S(\mathbf{q})$ are defined as~\cite{HansenMcDonald2013}
\begin{equation}
\rho g(\mathbf{r})=\frac{1}{N}\sum_i\sum_{j\ne i}
\left\langle \delta(\mathbf{r}+\mathbf{r}_i-\mathbf{r}_j)\right\rangle,
\end{equation}

\noindent
where $\rho$ is the particle density, and

\begin{equation}
S(\mathbf{q})=\frac{1}{N}\sum_{j=1}^N\sum_{l=1}^N
\left\langle e^{-i\mathbf{q}\cdot(\mathbf{r}_j-\mathbf{r}_l)}\right\rangle \quad .
\end{equation}
For an isotropic two-dimensional system, they depend only on $r=|\mathbf{r}|$ and $q=|\mathbf{q}|$, and are related by

\begin{equation}
S(q)=1+2\pi\rho\int_0^\infty r\,[g(r)-1]\,J_0(qr)\,dr,
\end{equation}
where \(J_0\) is the zeroth-order Bessel function of the first kind. 

As shown in Fig.~\ref{fig_gr_sq}, both increasing density and increasing roughness significantly affect the translational structure by modifying the peak positions and amplitudes in $g(r)$ and $S(q)$.

\subsection{The probability distribution of $\psi_6$}

To quantify the degree of local sixfold (hexatic) order discussed in the main text, we use the bond--orientational order parameter $\psi_6$. 

For a particle $k$, we first identify its nearest neighbors, here defined via the Voronoi construction. 

Denoting by $\theta_{kj}$ the angle of the bond vector from particle $k$ to a neighbor $j$ with respect to a fixed reference axis, the local sixfold bond-orientational order parameter is given by Eq.~(1) of the main text, i.e., 
\begin{equation}
\psi_6(k)=\left|\frac{1}{N_k}\sum_{j=1}^{N_k}\exp\!\left[\,\mathrm{i}\,6\theta_{kj}\right]\right| ,
\label{eq:q6_local}
\end{equation}

\noindent
where the sum runs over the $N_k$ Voronoi neighbors of particle $k$. Thus, $\psi_6(k)$ is obtained by averaging the phase factors associated with the bonds connecting particle $k$ to its neighbors.
 
By construction, $\psi_6(k)\in[0,1]$, with values close to unity indicating that the local environment of particle $k$ is consistent with a sixfold symmetric arrangement, while small values correspond to more disordered local neighborhoods. 
To characterize the structural heterogeneity of the system, we consider the distribution of $\psi_6$,

\begin{equation}
P(\psi)=\left\langle \delta\!\left(\psi-\psi_6\right)\right\rangle .
\label{eq:Pq6_def}
\end{equation}

\noindent

In Fig.~\ref{fig_P_q6_1} we show, for $\kappa=1.0$, the probability distribution $P(\psi_6)$ for different roughness values $R_a$ and various densities, which allows us to track how the population of locally ordered environments evolves with the control parameters.

\begin{figure}[ht]
\centering
\includegraphics[width=0.8\textwidth]{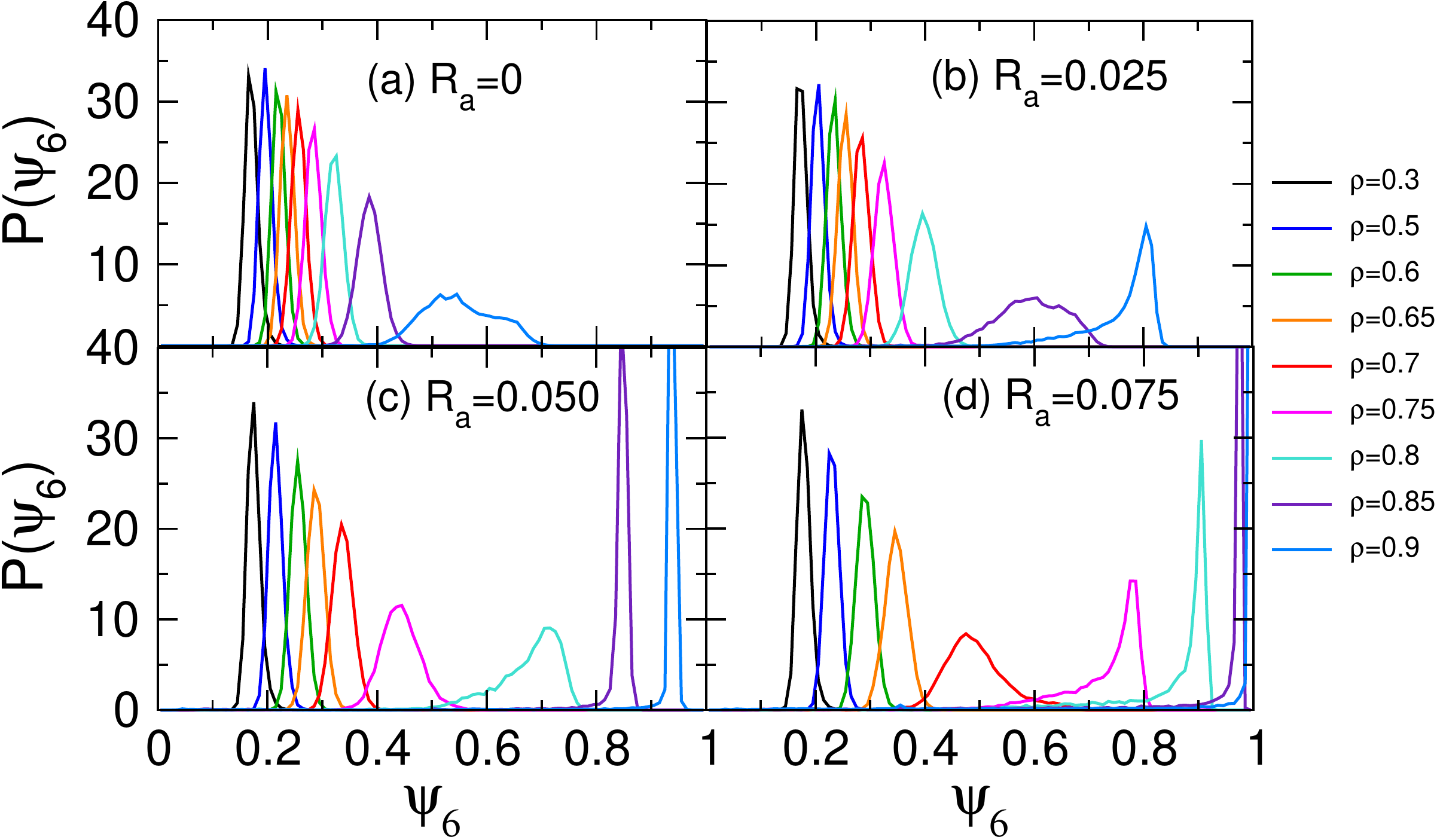}
\caption{
Probability distribution function $P(\psi_6)$ of the local hexatic order parameter $\psi_6$ for $\kappa=1.0$ at different number densities $\rho$ (color coded, from low to high $\rho$). 
Panels (a)--(d) correspond to roughness values $R_a=0$, $0.025$, $0.050$, and $0.075$, respectively. 
Upon increasing $\rho$, the distributions shift towards larger $\psi_6$ and progressively develop weight at high $\psi_6$, indicating the buildup of local sixfold order.\\ 
}

\label{fig_P_q6_1}
\end{figure}

\begin{figure}[ht]
\centering
\includegraphics[width=0.8\textwidth]{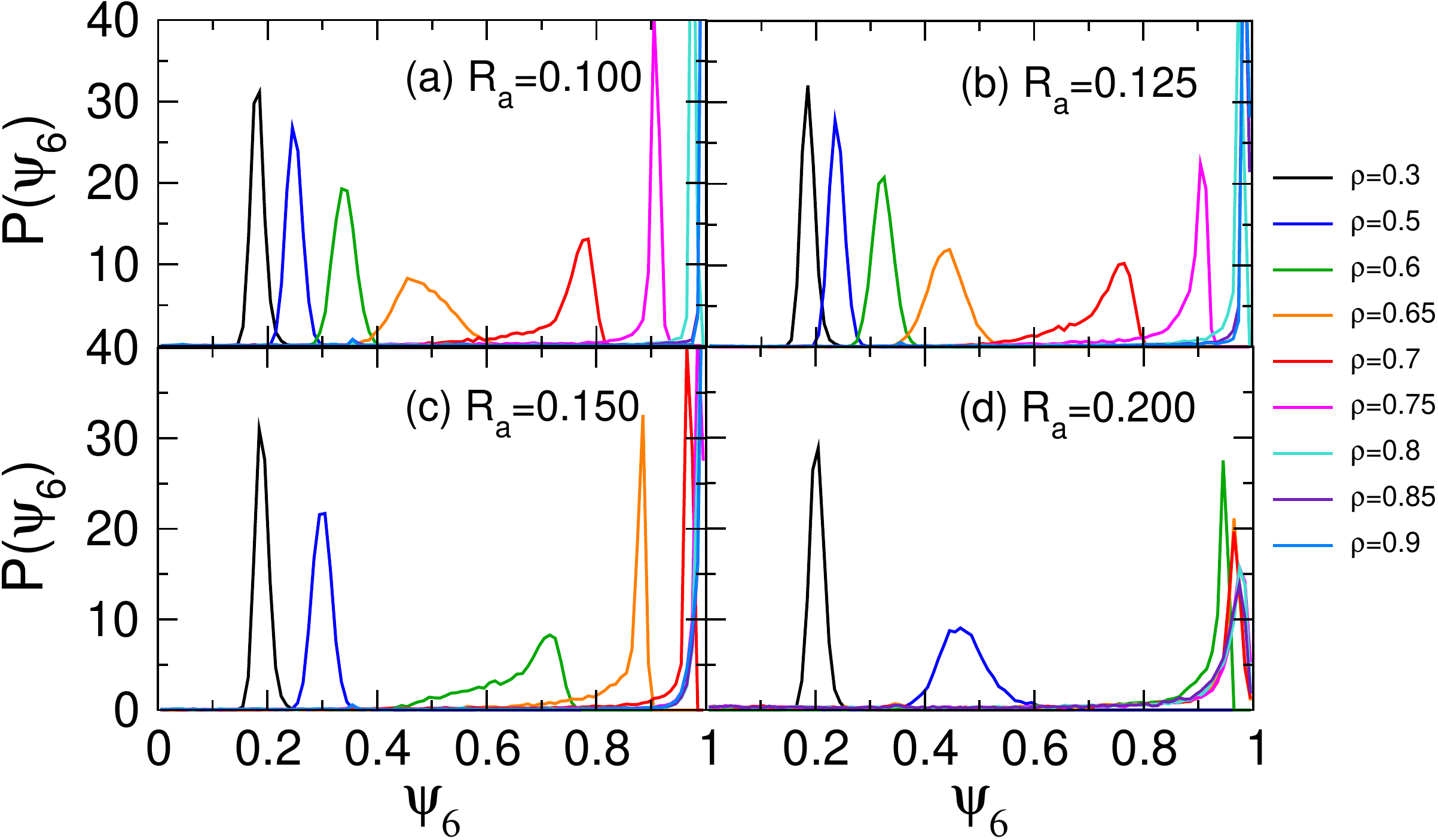}
\caption{Larger roughness values: (a) $R_a=0.100$, (b) $R_a=0.125$, (c) $R_a=0.150$, and (d) $R_a=0.200$ ($\kappa=1.0$). Different colors correspond to different number densities $\rho$. 
}
\label{fig_P_q6_2}
\end{figure}

With increasing density, the distributions $P(\psi_6)$ shift toward larger values of $\psi_6$, indicating an increasing degree of local sixfold order. This overall trend is similar for all values of $R_a$ shown. The main effect of roughness is quantitative: For larger $R_a$, the high-$\psi_6$ part of the distribution becomes more pronounced already at lower densities, and the high-density distributions are more strongly concentrated near $\psi_6 \approx 1$.

\subsection{Bond Angle Distribution}

Insight into the local order can be obtained from the bond-angle distribution function. 
To this end, we consider a central particle \(i\) and all pairs of its nearest neighbors \(j\) and \(k\). 
Nearest neighbors are identified using a cutoff radius \(r_c\), taken as the first minimum of the radial distribution function \(g(r)\). 
For each triplet \((j,i,k)\), the bond angle is defined as the angle between the two neighbor vectors connecting the central particle \(i\) to particles \(j\) and \(k\),

\begin{equation}
\theta_{jik}
= \arccos\!\left(
\frac{\mathbf{r}_{ij}\cdot\mathbf{r}_{ik}}
     {\|\mathbf{r}_{ij}\|\;\|\mathbf{r}_{ik}\|}
\right),
\label{eq_BAD}
\end{equation}
where \(\mathbf{r}_{ij} = \mathbf{r}_j - \mathbf{r}_i\) and 
\(\mathbf{r}_{ik} = \mathbf{r}_k - \mathbf{r}_i\).

\begin{figure}[th]
\centering
\includegraphics[width=1.0\textwidth]{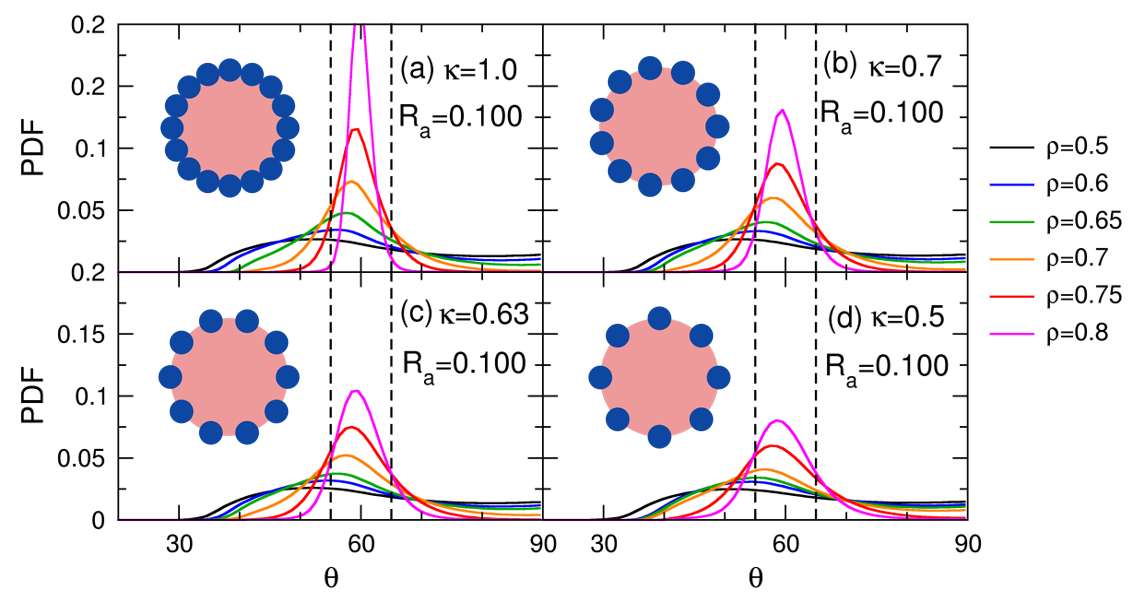}
\caption{Bond Angle distribution \(P(\theta)\) for the systems with coverage parameters $\kappa = 1.0$, $0.7$, $0.63$ and $0.5$, $R_a=0.1$ (panels (a)-(d)). The vertical dashed lines mark \(\theta=55^\circ\) and \(65^\circ\), i.e., the angular interval used to define \(A(\rho)\), the integral over the bond-angle distribution.
}
\label{fig_P_theta}
\end{figure}

Figure~\ref{fig_P_theta} shows the normalized bond-angle distribution for $R_a = 0.1$ at different values of $\kappa$. As the density increases, the peak near $60^\circ$ becomes progressively sharper, indicating the growing importance of locally triangular arrangements. To quantify this trend, we define the integrated bond-angle distribution $A(\rho)$ by integrating $P(\theta)$ over the interval $55^\circ \le \theta \le 65^\circ$,

\begin{equation}
A(\rho)=\int_{55^\circ}^{65^\circ} P(\theta;\rho)\, d\theta \, ,
\end{equation}

\noindent
which quantifies the number of bond angles in the vicinity of $60^\circ$. The two vertical dashed lines in each panel mark the lower and upper bounds of this integration window. 

Following the same procedure as for $\bar{\psi}_6$ in the main text, we then rescale the density axis to obtain a master-curve for each value of $\kappa$. From this collapse we determine the corresponding parameter $B(\kappa)$, which is then used to define the effective packing fraction $\phi_{\rm eff}$. The resulting $A(\phi_{\rm eff})$ is shown in Fig.~\ref{fig_A_phi}, where one sees that the data for the different values of $R_a$ fall nicely onto a master-curve when plotted as a function of $\phi_{\rm eff}$, in agreement with the result for $\bar{\psi}_6$ presented in Fig.~2 of the main text.

\begin{figure}[ht]
\centering
\includegraphics[width=0.8\textwidth]{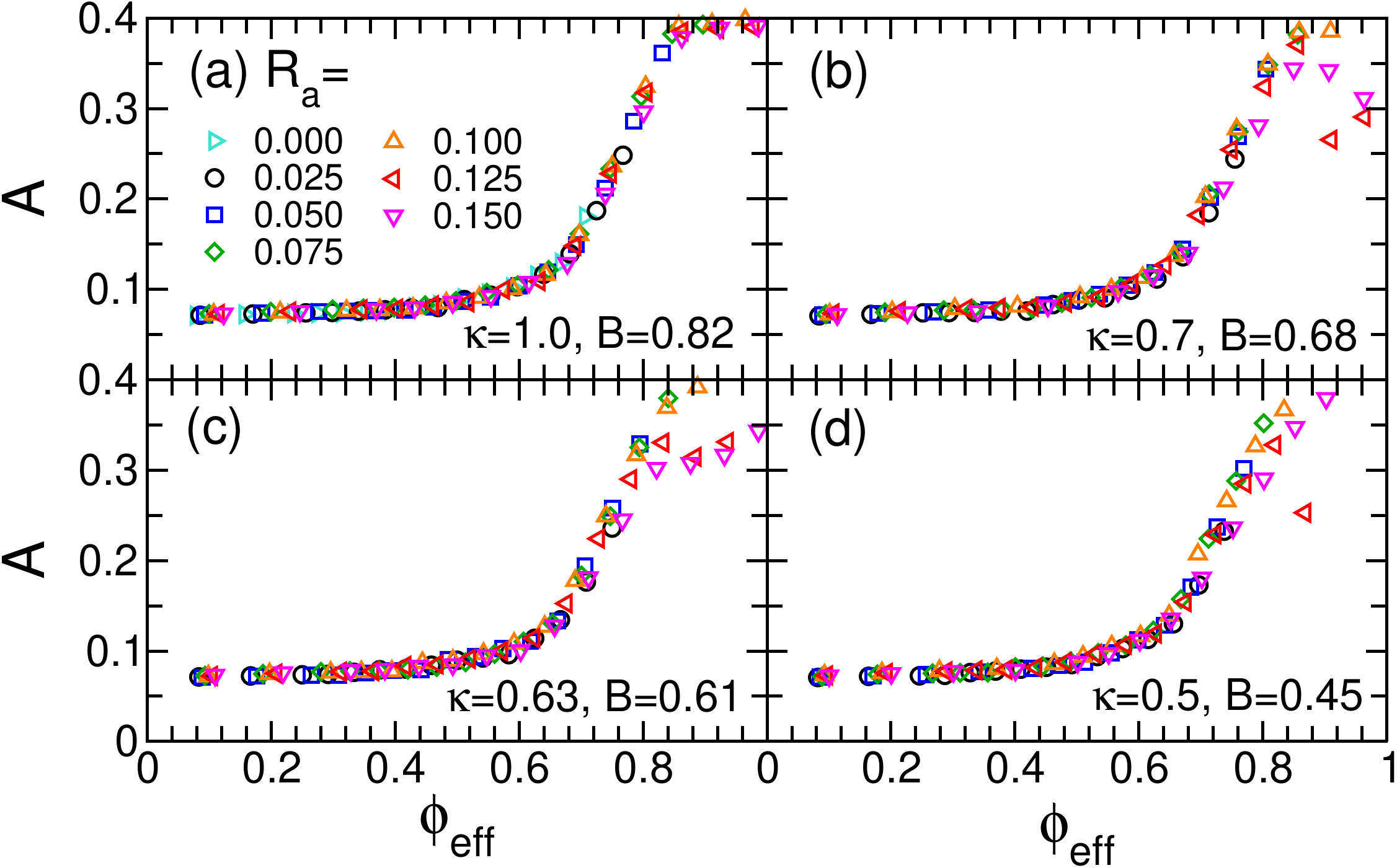}
\caption{
Integrated bond--angle distribution $A$ as a function of the effective packing fraction $\phi_{\mathrm{eff}}$ for different coverage parameters: (a) $\kappa=1.0$, (b) $\kappa=0.7$, (c) $\kappa=0.63$, and (d) $\kappa=0.5$. 
Different symbols correspond to the roughness values $R_a$ indicated in panel (a). 
For each $\kappa$, the data for different $R_a$ collapse onto a master curve when plotted versus $\phi_{\mathrm{eff}}$; the corresponding scaling factors $B$ are reported in each panel. 
}

\label{fig_A_phi}
\end{figure}

\subsection{Lindemann Ratio}

\begin{figure}[ht]
\centering
\includegraphics[width=0.8\textwidth]{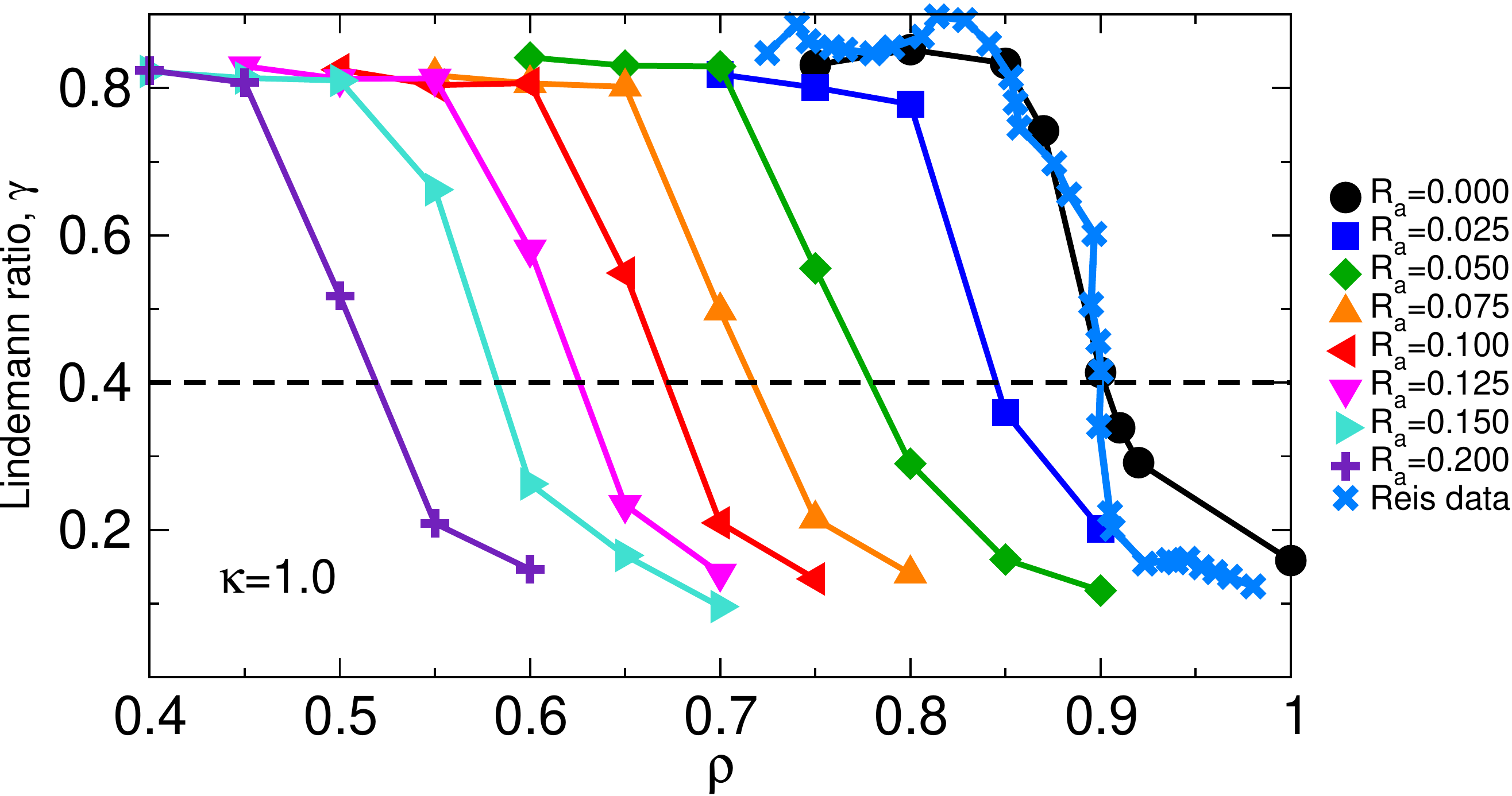}
\caption{Lindemann ratio $\gamma$ as a function of number density $\rho$ for particles with different surface roughness. The horizontal dashed line at $\gamma=0.4$ is used to determine the characteristic density $\rho_c$ for each roughness. Blue crosses indicate the experimental data reported by Reis {\it et al.}~\cite{Reis2006}. 
}
\label{fig_LR_gamma}
\end{figure}

To quantify the onset of crystallization based on positional disorder, we compute the Lindemann ratio $\gamma$ \cite{Reis2006}, defined as the root mean squared displacement of each particle relative to its time-averaged position, normalized by the average distance $L$ to its Voronoi neighbors: 

\begin{equation}
\gamma = \frac{1}{N} \sum_{i=1}^N \frac{\sqrt{\langle |\mathbf{r}_i - \langle \mathbf{r}_i \rangle_{T_{\rm win}}|^2 \rangle_{T_{\rm win}}}}{L} \quad .
\label{eq_LR}
\end{equation}

Here, $\langle\cdots\rangle_{T_{\rm win}}$ denotes a time average over the  window used to evaluate the Lindemann ratio. 

Specifically, we average over $M=1000$ configurations, sampled every $\Delta t$, so that the total window length is $T_{\rm win}=M\Delta t$.  
The window starts from the first recorded frame and ends at $t^\ast$, defined as the earliest time at which the mean-squared displacement (MSD) has entered the diffusive regime. To make this criterion robust, we determine $t^\ast$
from the instantaneous logarithmic slope $\alpha(t)\equiv d\ln \mathrm{MSD}(t)/d\ln t$ as the first time where $\alpha(t)\ge \alpha_{\rm th}$ with $\alpha_{\rm th}=0.95$.

The normalization length $L(R_a,\rho)$ is chosen as the (global) mean bond length between Voronoi nearest neighbors, i.e., the average distance between a particle and its Voronoi neighbors, which provides an estimate of the typical local spacing in the dense phase~\cite{Reis2006}. 

Figure~\ref{fig_LR_gamma} shows the evolution of the Lindemann ratio as a function of density. For all roughness values, the Lindemann ratio drops sharply as the system approaches the crystallization transition, and different roughnesses exhibit nearly identical trends. Therefore, the fitting procedure described in the main text for $\bar{\psi}_6$ can be applied to extract the scaling factor $B$. Also included is the experimental data of Reis et al.\cite{Reis2006} for a two-dimensional (smooth) hard-disk system. One sees that this data agrees qualitatively with our $R_a=0$ curve, but that the drop is steeper. A plausible reason for this difference is that the experimental particles are steel spheres, whose effective normal stiffness is roughly $10^{3}$ times larger than the one in our model, leading to a faster stiffening of the cages and thus a sharper drop of the Lindemann ratio.

\subsection{Triplet Local Order Parameter $T_6$}

While \(\psi_6\) characterizes sixfold bond-orientational order through the orientations of the bonds connecting a particle to its neighbors, it is useful to complement it by a quantity that probes the local angular structure more directly. For this purpose, we introduce the triplet-based orientational parameter \(T_6\), which is constructed from the bond angles formed by mutually connected neighboring triplets and is therefore particularly sensitive to short-range triangular order. 

Specifically, for a given central particle $p$, we order the relevant Voronoi neighbors around $p$ consecutively and denote their number by $N_T(p)$. For each pair of successive neighbors $j$ and $j+1$, we define the angle $\theta_{jp(j+1)}$ at particle $p$. The local order parameter is then given by

\begin{equation}
T_6^{(p)}=
\left|
\frac{1}{N_T(p)}
\sum_{j=1}^{N_T(p)}
\exp\!\left(i\,6\,\theta_{jp(j+1)}\right)
\right| \quad .
\label{eq_T6}
\end{equation}

The global $T_6$ is then defined as
\begin{equation}
T_6=\left\langle \frac{1}{N}\sum_{p=1}^{N} T_6^{(p)} \right\rangle.
\label{eq_T6_global}
\end{equation}

The parameter $T_6$ quantifies the degree of sixfold angular symmetry in the local neighborhood~\cite{Tong2018HiddenOrder}. As for the other structural observables discussed above, we rescale the density axis for each value of $\kappa$ in order to collapse the data for different roughnesses $R_a$ onto a single master curve, from which we obtain the corresponding factor $B$ and hence the effective packing fraction $\phi_{\rm eff}$, see Fig.~\ref{fig_w6_phi}.

\begin{figure}[ht]
\centering
\includegraphics[width=0.8\textwidth]{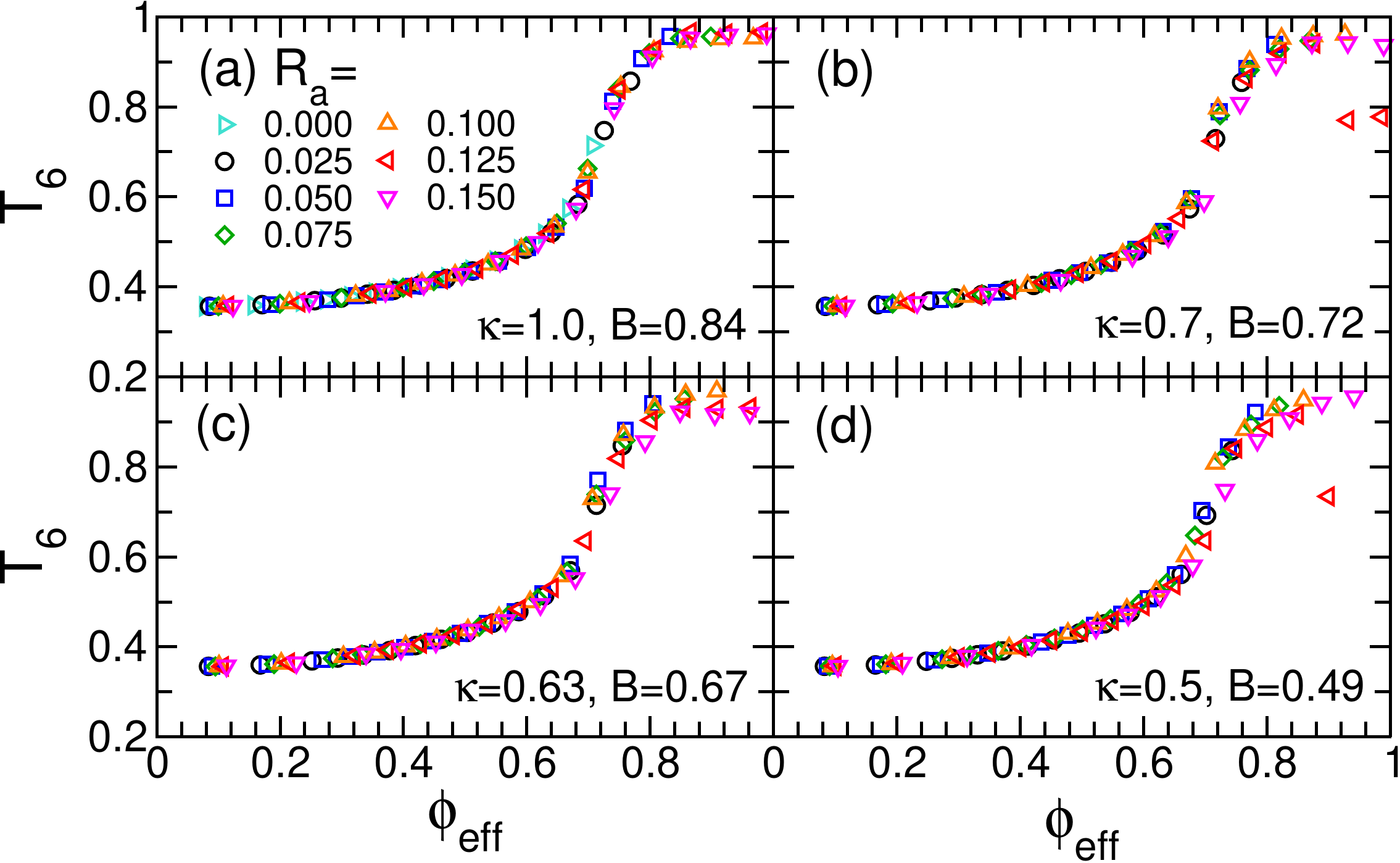}
\caption{Collapse of the local order parameter $T_6$ versus the effective packing fraction $\phi_{\rm eff}$ for different roughness values $R_a$. Panels (a)--(d) show different coverages $\kappa$, with the corresponding rescaling factor $B$ indicated. 
} 
\label{fig_w6_phi}
\end{figure}

At fixed coverage $\kappa$, the $T_6$ data for different roughness values $R_a$ collapse onto a single master curve when plotted against $\phi_{\rm eff}$, using a rescaling factor $B$ obtained for this value of $\kappa$.
This indicates that the growth of local sixfold angular order is primarily controlled by $\phi_{\rm eff}$ and that the effect of roughness can be largely absorbed by the geometric rescaling.

\subsection{Mean squared displacement and cage-relative mean squared displacement}

\begin{figure}[ht]
\centering
\includegraphics[width=0.8\textwidth]{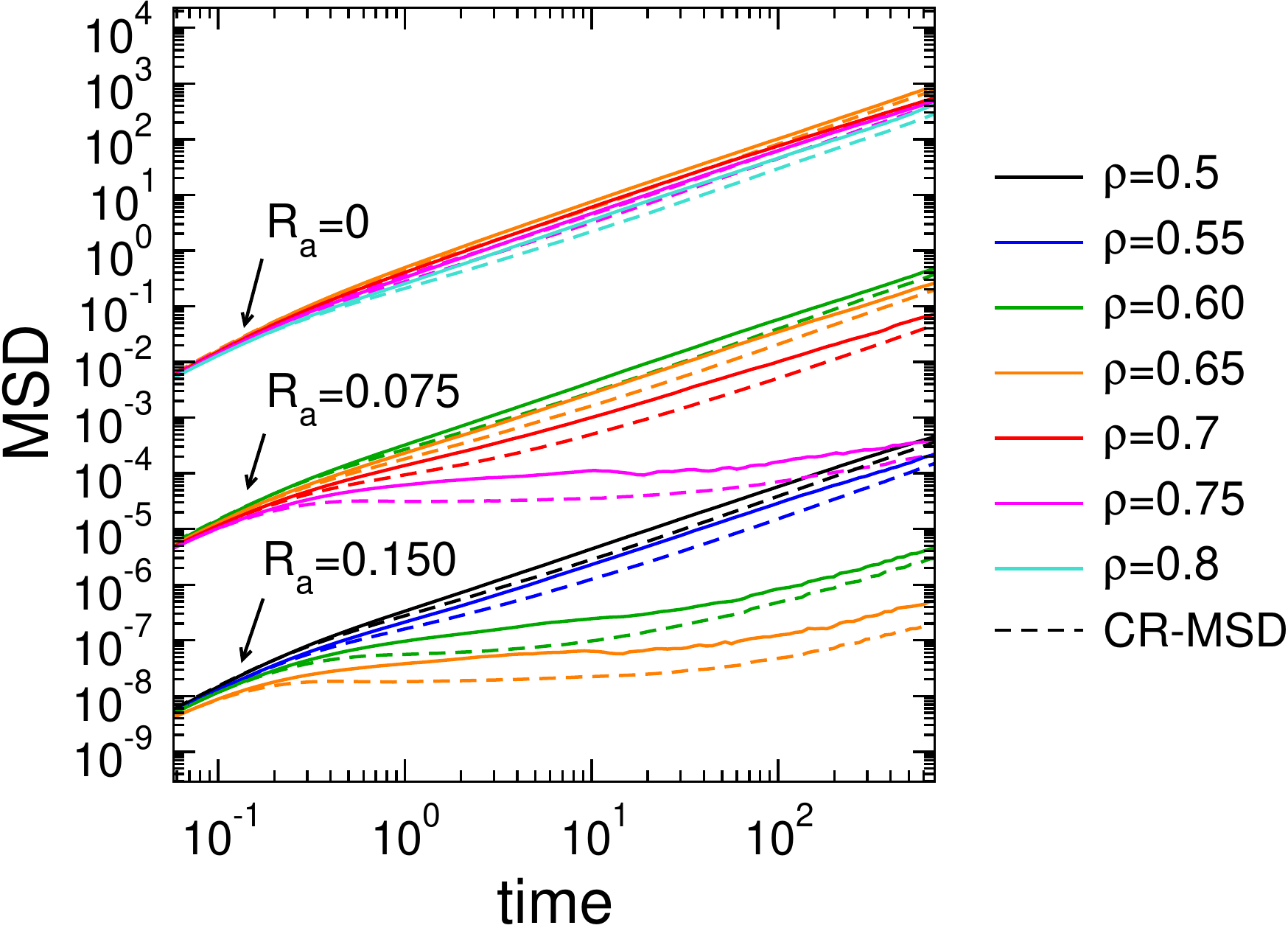}
\caption{Different colors correspond to different number densities. The solid lines show the MSD (mean-squared displacement), while the dashed lines represent the CR-MSD (cage-relative MSD). For clarity, the results for $R_a = 0.075$ and $R_a = 0.150$ are divided by $10^{3}$ and $10^{6}$, respectively, to separate the data for the three roughness values. All results shown here correspond to $\kappa = 1.0$.
}
\label{fig_kap1.0_msd_CR-msd}
\end{figure}

We calculate the mean squared displacement (MSD), which is defined as~\cite{frenkel2002}
\begin{equation}
\left\langle r^{2}(t)\right\rangle = \frac{1}{N} \sum_{j=1}^{N}  \langle \left[\vec{r}_{j}(t) - \vec{r}_{j}(0)\right]^{2} \rangle,
\label{eq_Tmsd}
\end{equation}

\noindent
where $t$ represents the time interval over which the displacement is measured. 

In large dense two-dimensional systems, the tagged particle dynamics is strongly affected by Mermin--Wagner fluctuations~\cite{flenner2015fundamental}. 
These collective modes lead to coherent motions on large length scales, which contribute substantially to the single-particle displacements. 
As a consequence, the standard MSD becomes strongly affected by such collective fluctuations and is therefore not a reliable probe of structural relaxation in 2D since the growth of the mean squared displacement (MSD) is dominated by these fluctuations rather than the true diffusive motion, i.e., the conventional MSD analysis becomes unreliable under high-density conditions. To remove this contribution, we consider the cage-relative mean squared displacement (CR-MSD), in which the displacement of each particle is measured relative to the average displacement of its neighbors~\cite{illing2017mermin}, defined as

\begin{equation}
\begin{aligned}
\left\langle r^{2}(t) \right\rangle^{\mathrm{CR}} 
= \frac{1}{N} \sum_{j=1}^{N} \Bigg\langle \Big[
\vec{r}_{j}(t) - \vec{r}_{j}(0)
-\frac{1}{N_{j}} \sum_{i=1}^{N_{j}} \big( \vec{r}_{i}(t) - \vec{r}_{i}(0) \big)
\Big]^2 \Bigg\rangle .
\end{aligned}
\label{eq:CR-MSD}
\end{equation}

\noindent
where the $i-$sume runs over the neighbors of particle $j$ at $t=0$ and $N_{j}$ is the number of neighbors of particle $j$. So $\left\langle r^{2}(t) \right\rangle^{\mathrm{CR}}$ is the cage-relative mean-squared displacement at lag time $t$.

As shown in Fig.~\ref{fig_kap1.0_msd_CR-msd}, both the MSD and the CR-MSD evolve towards a long-time diffusive regime with slope 1. (In the full time window one also observes the short-time ballistic regime with slope 2 in this double-logarithmic representation; however, the figure shown here is zoomed on the intermediate- and long-time window, and hence this ballistic part is not visible.) At intermediate times, the MSD is noticeably larger than the CR-MSD, reflecting the contribution of collective long-wavelength fluctuations to the single-particle motion. 
While the MSD and CR-MSD need not coincide at long times in two dimensions, we find that the corresponding diffusion coefficients $D$ and $D_{\rm CR}$ remain very similar. As shown in Fig.~\ref{fig_kap1.0_D_D_CR}, the diffusion coefficients extracted from the MSD and from the CR-MSD, $D$ and $D_{\rm CR}$, track each other closely over the explored density range. This suggests that, for the system sizes studied here, long-wavelength (Mermin--Wagner) fluctuations do not significantly affect the determination of $D$, so that using the standard MSD is sufficient to characterize the dyanmics.

\begin{figure}[ht]
\centering
\includegraphics[width=1.0\textwidth]{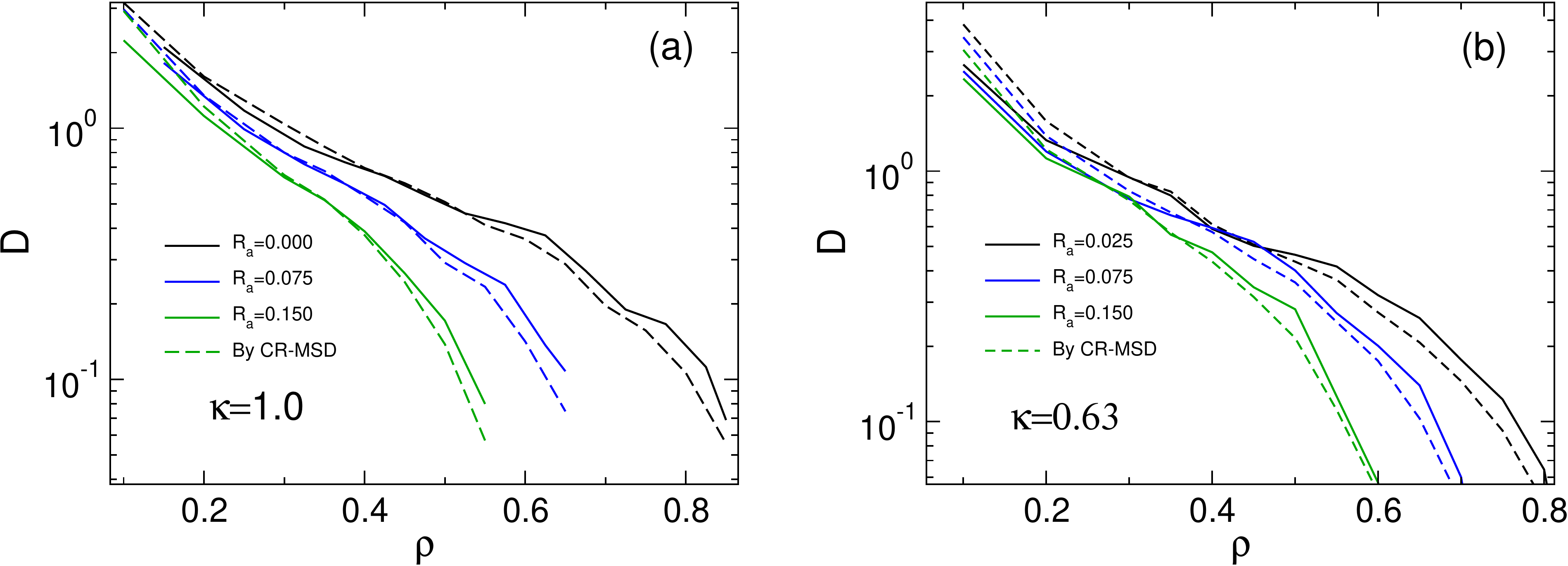}
\caption{(a) and (b) are the results of the diffusion coefficients $D$ with $\kappa=1.0$ and $\kappa=0.63$, respectively. The solid line is obtained through the calculation of MSD, and the dashed line is obtained through the calculation of CR-MSD.}
\label{fig_kap1.0_D_D_CR}
\end{figure}

\clearpage
\subsection{The analytical derivation of B}

In this section we present  the details of the calculations that allow to estimate the value of the parameter $B$ defined in Eq.~(2) of the main text, i.e.,

\begin{equation}
R_{\mathrm{eff}} = R_c\left(1 + B\,\frac{R_a}{R_c}\right).
\end{equation}

Thus $B$ quantifies how much $R_a$ affects the effective radius $R_{\rm eff}$ of a MP. We assume that $R_{\rm eff}$ is directly proportional to the distance between two MPs that touch each other. In the following we will thus express this distance as a function of $R_a$ and $\kappa$ and then consider the limit that $R_a/R_c$ is small, which results in an explicit expression for $B$.

Figure~\ref{fig:B7-1} shows a typical configuration for two MPs that touch each other. The distance $d_{\rm OC}$ between the two centers depends on the relative orientation of the two MPs, i.e., on the angles $\delta$ and $\beta$ defined in the figure. Thus to determine the mean distance $\bar{d}$ between the two centers, one has to take the average over these two angles. Figures~\ref{fig:B7-2} and \ref{fig:B7-3} show that to prevent overlap between the MPs the angle $\beta$ is restricted by an upper and lower bound, $\beta_{min}$ and $\beta_{\rm max}$, respectively and the same holds for the angle $\delta$, thus defining $\delta_{\rm min}$ and $\delta_{\rm max}$.

\begin{figure}[h!]
    \centering
    \includegraphics[width=0.8\textwidth]{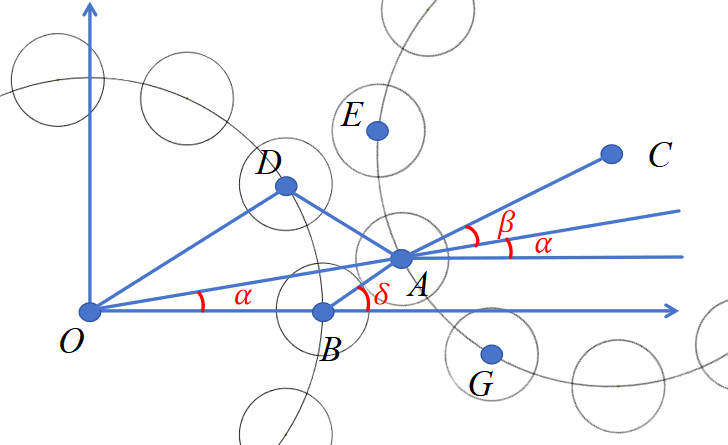}
    \caption{Typical contact configuration between two MPs and the definition of various angles used in the presented calculation.
}
    \label{fig:B7-1}
\end{figure}

\vspace*{20mm}

\begin{figure}[h!]
    \centering
    \includegraphics[width=0.8\textwidth]{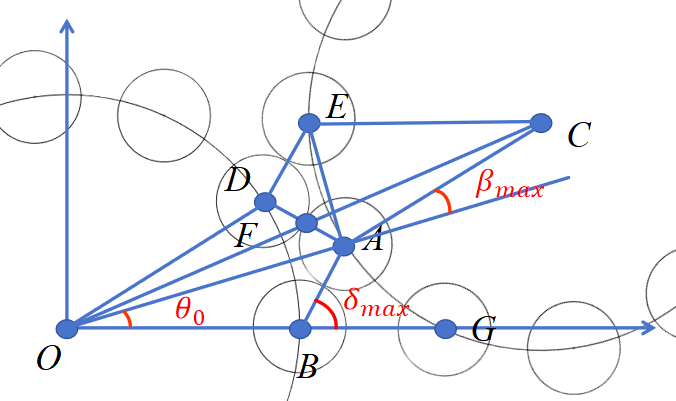}
    \caption{Configuration in which the right MP is rotated upwards up to its maximum extent, thus defining $\delta_{\max}$ and $\beta_{\max}$. The angle $\theta_0$ denotes half of the angular interval between two neighboring asperities on the same MP. 
    }
    \label{fig:B7-2}
\end{figure}

\begin{figure}[h!]
    \centering
    \includegraphics[width=0.8\textwidth]{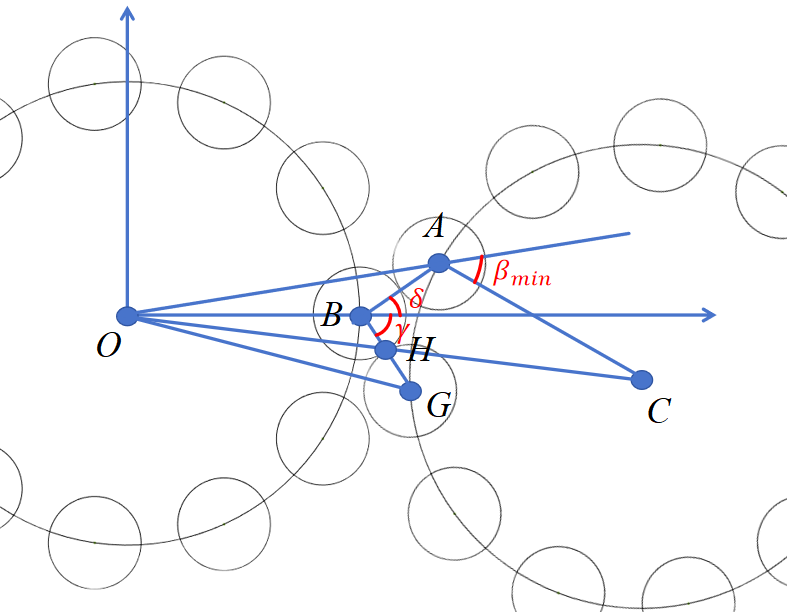}
    \caption{Configuration in which for a given angle $\delta$ the right MP is rotated downwards up to the minimum angle $\beta_{\rm min}$.
    }
    \label{fig:B7-3}
\end{figure}

By placing in Fig.~\ref{fig:B7-1} the origin of the coordinate system at the center of the left MP and orienting the $x$-axis along the connection between this center and one of the asperities, we can read off the coordinates/values of the following points/angles :  

\begin{equation}
\mathrm{A}:(R_c+2R_a\cos\delta,2R_a\sin\delta),\qquad \mathrm{C}:(R_c+2R_a\cos\delta+R_c\cos(\alpha+\beta),2R_a\sin\delta+R_c\sin(\alpha+\beta))
\label{eq:coord_AC}
\end{equation}
\begin{equation}
\tan\alpha=\frac{2R_a\sin\delta}{R_c+2R_a\cos\delta}
\end{equation}

So the distance $d_{\rm OC}$ between the two centers is given by the location of point C and is given by: 
\begin{align}
\bar{d}_{\mathrm{OC}}
&= \frac{1}{2\delta_{\max}\,(\beta_{\max}-\beta_{\min})}
   \int_{-\delta_{\max}}^{\delta_{\max}}
   \int_{\beta_{\min}}^{\beta_{\max}} |{\bf OC}(\beta,\delta)| \, d\beta\, d\delta \quad .
\label{eq:numerical_original}
\end{align}

In the following we will derive exact expressions for $\delta_{\min}$, $\delta_{\max}$, $\beta_{\min}$, and $\beta_{\max}$, and subsequently we will return to the expression (\ref{eq:numerical_original}) for calculating the average distance between two MPs that touch each other.

The angle $\delta_{\max}$ can be obtained by evaluating the $\angle OBA$, as shown in Fig.~\ref{fig:B7-2}. For this we introduce the angle $\theta_0$, which is half of the angular interval between two neighboring asperities on the same MP (see Fig.~\ref{fig:B7-2}). Since the $N_a$ asperities are uniformly distributed along the circumference of the central disk, one has $\theta_0=2\pi/(2N_a)$. Furthermore, using $\kappa=N_aR_a/(\pi R_c)$, it follows that $\theta_0=(R_a/R_c)/\kappa=x/\kappa$, where $x=R_a/R_c$.

\begin{align}
\lvert\mathbf{OA}\rvert & {=} R_c\cos\theta_0+ \sqrt{(2R_a)^2 - (R_c\sin\theta_0)^2}, 
\label{eq:norm_OA_2}
\end{align}

This gives for the angle
\begin{equation}
\cos\angle OBA {=} \frac{4R_a^2+R_c^2-(R_c\cos\theta_0+\sqrt{4R_a^2-R_c^2\sin\theta_0^2})^2}{4R_aR_c} \quad .
\label{eq:cosdelta_expand}
\end{equation}

which can be rewritten as
\begin{equation}
\cos\angle OBA =\frac{\sin^2\theta_0-\cos\theta_0\sqrt{4x^2-\sin^2\theta_0}}{2x} \quad.
\label{eq:cos_delta_nondim}
\end{equation}

So we obtain for $\delta_{max}$ (see Fig.~\ref{fig:B7-2}):

\begin{equation}
\boxed{\delta_{\max}=\arccos\!\left(\frac{\cos\theta_0\sqrt{4x^2-\sin^2\theta_0}
-\sin^2\theta_0}{2x}\right)}
\label{eq:numerical_delta_max}
\end{equation}

This is an exact expression. Because the MPs are symmetric with respect to the OB axis, we have {$\delta_{\min} = -\delta_{\max}$}.

We can simplify the formula by making the assumption that $x \ll 1$. For this, we expand the trigonometric functions as
\begin{equation}
\sin(x/\kappa) \approx \frac{x}{\kappa},
\qquad
\cos(x/\kappa) \approx 1 - \frac{x^2}{2\kappa^2}.
\label{eq:trig_expand}
\end{equation}

Substituting Eq.~\eqref{eq:trig_expand} into Eq.~\eqref{eq:cos_delta_nondim}, we obtain
\begin{align}
\cos\angle OBA&\approx\frac{\frac{x^2}{\kappa^2}-(1-\frac{x^2}{2\kappa^2})\sqrt{4x^2-\frac{x^2}{\kappa^2}}}{2x}, 
\label{eq:cos_delta_expand}
\end{align}

which can be rewritten as
\begin{align}
\cos\angle OBA
&\approx-\frac{1}{2}\sqrt{4-\frac{1}{\kappa^2}}+\frac{x}{2\kappa^2}+\frac{x^2\sqrt{4-\frac{1}{\kappa^2}}}{4\kappa^2} \quad .
\label{eq:cos_delta_simplified}
\end{align}

In the limit $x \to 0$, the leading term is thus
\begin{equation}
\cos\angle OBA
\;\approx\; 
-\frac{1}{2}\sqrt{4 - \frac{1}{\kappa^2}}=-\sqrt{1-\frac{1}{4\kappa^2}}.
\label{eq:cos_delta_limit}
\end{equation}

\noindent
resulting in 

\begin{equation}
\boxed{\delta_{\max} {\approx} \arccos(\sqrt{1-\frac{1}{4\kappa^2}})}
\label{eq_delta_max_approx}
\end{equation}

Now we determine $\beta_{\max} $ for a generic case. Figure~\ref{fig:B7-2} shows that we have:
\begin{align}
    \mathrm{A}:(R_c+2R_a\cos\delta,2R_a\sin\delta),\qquad \mathrm{D}:(R_c\cos2\theta_0,R_c\sin2\theta_0)
\end{align}

Because the $|{\bf DE}|=|{\bf AB}|$ and $|{\bf BD}|={\bf AE}|$, the quadrilateral $ABDE$ is a parallelogram.
So the coordinates of E can be written as:
\begin{equation}
\mathrm{E}: \; \bigl(R_c\cos 2\theta_0 + 2R_a\cos\delta,\;
          R_c\sin 2\theta_0 + 2R_a\sin\delta\bigr).
\end{equation}

Point $C$ is located at the intersection of two circles centered at $A$ and $E$,  
each having the same radius $R_c$, i.e.,
\begin{align}
(x - x_A)^2 + (y - y_A)^2 &= R_c^2, \label{eq_circle_A}\\
(x - x_E)^2 + (y - y_E)^2 &= R_c^2. \label{eq_circle_E}
\end{align}

Subtracting Eq.~\eqref{eq_circle_E} from Eq.~\eqref{eq_circle_A}, gives:
\begin{equation}
(x_E - x_A)x + (y_E - y_A)y
= \frac{1}{2}\bigl(x_E^2 - x_A^2 + y_E^2 - y_A^2\bigr).
\label{eq_chord_line}
\end{equation}

Let the distance between $A$ and $E$ be
\begin{equation}
f = 2R_c\sin\theta_0,
\label{eq_d_def}
\end{equation}
and define $h$ as the perpendicular offset from the midpoint of $AE$ to the intersection point $C$:
\begin{equation}
h = \sqrt{R_c^2 - \frac{f^2}{4}}=R_c\cos\theta_0.
\label{eq_h_def}
\end{equation}

Then the coordinates of $C$ (for the intersection on the right-hand side of $AE$) are given by the standard
two-circle intersection formula:
\begin{align}
C_x &= \frac{x_A + x_E}{2} + \frac{y_E - y_A}{d}\,h, \label{eq_Cx_general}\\[3pt]
C_y &= \frac{y_A + y_E}{2} - \frac{x_E - x_A}{d}\,h. \label{eq_Cy_general}
\end{align}

The midpoint of ${\bf AE}$ is
\begin{align}
\frac{x_A + x_E}{2} &= R_c\cos^2\theta_0 + 2R_a\cos\delta, \\
\frac{y_A + y_E}{2} &= R_c\sin\theta_0\cos\theta_0 + 2R_a\sin\delta.
\end{align}
From these expressions we thus obtain
\begin{align}
C_x &= R_c\cos^2\theta_0 + 2R_a\cos\delta
     + \frac{R_c\sin 2\theta_0}{2R_c\sin\theta_0}\,R_c\cos\theta_0
     = R_c(1+\cos 2\theta_0) + 2R_a\cos\delta, \label{eq_Cx_final}\\[4pt]
C_y &= R_c\sin\theta_0\cos\theta_0 + 2R_a\sin\delta
     - \frac{R_c(\cos 2\theta_0 - 1)}{2R_c\sin\theta_0}\,R_c\cos\theta_0
     = R_c\sin 2\theta_0 + 2R_a\sin\delta. \label{eq_Cy_final}
\end{align}

Therefore, the coordinates of point $C$ are given by
\begin{equation}
\mathrm{C}:\bigl(R_c(1+\cos 2\theta_0) + 2R_a\cos\delta,\;
       R_c\sin 2\theta_0 + 2R_a\sin\delta\bigr).
\label{eq_C_coord}
\end{equation}

The corresponding vectors are
\[
\mathbf{OA} = (R_c + 2R_a\cos\delta,\; 2R_a\sin\delta), \qquad
\mathbf{AC} = (R_c\cos 2\theta_0,\; R_c\sin 2\theta_0).
\]

The cosine of the angle $\beta_{\max}$ is given by
\begin{equation}
\cos\beta_{\max}
= \frac{{\bf OA}\cdot {\bf AC}}{|{\bf OA}||{\bf AC}|}.
\label{eq_cos_OAAC_def}
\end{equation}
The dot product reads
\begin{align}
\mathbf{OA}\cdot\mathbf{AC}
&= (R_c + 2R_a\cos\delta)(R_c\cos 2\theta_0)
 + (2R_a\sin\delta)(R_c\sin 2\theta_0) \notag\\[3pt]
&= R_c^2\cos 2\theta_0
 + 2R_aR_c(\cos\delta\cos 2\theta_0 + \sin\delta\sin 2\theta_0).
\label{eq_dot_OAAC}
\end{align}
The vector magnitudes are
\begin{align}
|\mathbf{OA}|
&= \sqrt{(R_c + 2R_a\cos\delta)^2 + (2R_a\sin\delta)^2}
 = \sqrt{R_c^2 + 4R_aR_c\cos\delta + 4R_a^2}, \label{eq_norm_OA}\\[3pt]
|\mathbf{AC}|
&= \sqrt{(R_c\cos 2\theta_0)^2 + (R_c\sin 2\theta_0)^2}
 = R_c. \label{eq_norm_AC}
\end{align}
Substituting Eqs.~\eqref{eq_dot_OAAC}--\eqref{eq_norm_AC} into
Eq.~\eqref{eq_cos_OAAC_def} gives
\begin{equation}
\cos\beta_{\max}=\frac{\cos 2\theta_0 + 2\frac{R_a}{R_c}(\cos\delta\cos 2\theta_0 + \sin\delta\sin 2\theta_0)}{\sqrt{1 + 4\frac{R_a}{R_c}\cos\delta + 4\bigl(\frac{R_a}{R_c}\bigr)^2}},
\label{eq_cos_OAAC_exact}
\end{equation}
resulting in
\begin{equation}
\boxed{\beta_{\max}=\arccos\!\left[\frac{\cos\!\left(\frac{2x}{\kappa}\right)+
2x\left(\cos\delta \cos\!\left(\frac{2x}{\kappa}\right)+\sin\delta\sin\!\left(\frac{2x}{\kappa}\right)\right)}{\sqrt{1+4x\cos\delta+4x^2}}\right]}
\label{eq:numerical_beta_max}
\end{equation}

This is an exact expression. 
For small $x=R_a/R_c\ll1$, the denominator can be expanded as
\[
\sqrt{1+4x\cos\delta+4x^2}\approx 1+2x\cos\delta.
\]
Retaining only terms up to first order in $x$, Eq.~\eqref{eq_cos_OAAC_exact} becomes
\begin{align}
\cos\beta
&\approx
\bigl[\cos 2\theta_0
 + 2x(\cos\delta\cos 2\theta_0 + \sin\delta\sin 2\theta_0)\bigr](1 - 2x\cos\delta) \notag\\[3pt]
&\approx
\cos 2\theta_0 + 2x\sin\delta\sin 2\theta_0.
\label{eq_cos_OAAC_expand}
\end{align}

We define $\varepsilon$ by means of $\beta = 2\theta_0 + \varepsilon$. Using 
$\cos(2\theta_0+\varepsilon)\approx\cos2\theta_0-\varepsilon\sin2\theta_0$, we thus find
\[
-\varepsilon\sin2\theta_0 = 2\frac{R_a}{R_c}\sin\delta\sin2\theta_0
\quad\Rightarrow\quad
\varepsilon = -\,2\frac{R_a}{R_c}\sin\delta.
\]
Therefore, to first order in $R_a/R_c$, we obtain
\begin{equation}
\boxed{\beta_{\max} \approx 2\theta_0 - 2\frac{R_a}{R_c}\sin\delta=2x(\frac{1}{\kappa}-\sin\delta).}
\label{eq_beta_max_approx}
\end{equation}

To determine the lower bound $\beta_{\min}$, we consider in figure \ref{fig:B7-3} the triangle $ABG$ and obtain
\begin{align}
\cos(\gamma+\delta) &= \frac{(2R_a)^2+(2R_a)^2-(2R_c\sin\theta_0)^2}{8R_a^2}\\
&=1-\frac{\sin^2\theta_0}{2x^2}\\
&=1-\frac{\sin^2\frac{x}{\kappa}}{2x^2}\\
&\approx1-\frac{1}{2\kappa^2} \quad .
\label{eq:cosdelat+gamma}
\end{align}

In a similar way we find

\begin{align}
\sin(\gamma+\delta) &=\sqrt{1 - \cos^2(\gamma+\delta)}\\
&\approx\frac{1}{\kappa}\sqrt{1-\frac{1}{4\kappa^2}} \quad .
\label{eq:sindelat+gamma}
\end{align}

Combining Eqs.~(\ref{eq:cosdelat+gamma}) and (\ref{eq:sindelat+gamma}) gives
\begin{align}
\sin\gamma & = \frac{1}{\kappa}\sqrt{1-\frac{1}{4\kappa^2}}\cos\delta - (1-\frac{1}{2\kappa^2})\sin\delta
\label{eq:singamma}
\end{align}

\begin{align}
\cos\gamma&= (1-\frac{1}{2\kappa^2})\cos\delta + \frac{1}{\kappa}\sqrt{1-\frac{1}{4\kappa^2}}\sin\delta \quad .
\label{eq:cosgamma}
\end{align}

Using the angle $\gamma$, the coordinates of the points $A, H,$ and $C$ can be expressed as:
\begin{equation}
\mathrm{A}:(R_c+2R_a\cos\delta,2R_a\sin\delta), \qquad 
\mathrm{H}:(R_c+R_a\cos\gamma,R_a\sin\gamma),\qquad \mathrm{C}:(2(R_c+R_a\cos\gamma),2R_a\sin\gamma),\qquad
\label{eq:AHC_points}
\end{equation}

We hence obtain
\begin{align}
\mathbf{OA} &= (R_c + 2R_a\cos\delta,\; 2R_a\sin\delta), \label{eq:OA_vec}\\
\mathbf{AC} &= \mathbf{C}-\mathbf{A}
= \bigl[R_c + 2R_a(\cos\gamma - \cos\delta),\;
2R_a(\sin\gamma - \sin\delta)\bigr]. \label{eq:AC_vec}
\end{align}
The internal angle at $A$ is
\begin{equation}
\cos\beta_{\min}=\frac{\mathbf{OA}\cdot\mathbf{AC}}
{|\mathbf{OA}|\,|\mathbf{AC}|}.
\label{eq:cosOAC_def}
\end{equation}
The scalar product reads
\begin{align}
\mathbf{OA}\cdot\mathbf{AC}
&=(R_c+2R_a\cos\delta)\,[R_c+2R_a(\cos\gamma-\cos\delta)]
+(2R_a\sin\delta)\,[2R_a(\sin\gamma-\sin\delta)] \notag\\
&= R_c^2 + 2R_cR_a\cos\gamma+ 4R_a^2[\cos(\gamma-\delta)-1].
\label{eq:dot_product}
\end{align}
The norms are
\begin{align}
|\mathbf{AO}| &= R_c\sqrt{1 + 4x\cos\delta + 4x^2}, \label{eq:norm_AO}\\
|\mathbf{AC}| &= R_c.
\label{eq:norm_AC}
\end{align}
Substituting Eqs.~\eqref{eq:dot_product}--\eqref{eq:norm_AC} into
Eq.~\eqref{eq:cosOAC_def} gives the expression
\begin{equation}
\cos\beta_{\min}{=}\frac{1 + 2x\cos\gamma+ 4x^2[\cos(\gamma-\delta)-1]}{\sqrt{1 + 4x\cos\delta + 4x^2}},
\label{eq:cosOAC_nondim}
\end{equation}

\noindent
which gives for $\beta_{\min}$
\begin{equation}
\boxed{\beta_{\min}=\arccos(\frac{1 + 2x\cos\gamma+ 4x^2[\cos(\gamma-\delta)-1]}{\sqrt{1 + 4x\cos\delta + 4x^2}})} \quad ,
\label{eq:numerical_beta_min}
\end{equation}

\noindent
where the angle $\gamma$ is given by Eqs.~(\ref{eq:singamma}) and (\ref{eq:cosgamma}).
This is an exact expression. To simplify it in the small-$x$ limit we Taylor expand the inverse of the square root,
\[
\frac{1}{\sqrt{1+\epsilon}}
= 1 - \frac{1}{2}\epsilon + \frac{3}{8}\epsilon^{2} + O(\epsilon^{3}),
\]
with $\epsilon = 4x\cos\delta + 4x^{2}$. Substituting this expression gives
\begin{align}
\frac{1}{\sqrt{1 + 4x\cos\delta + 4x^{2}}}
&\approx 1 - 2x\cos\delta - 2x^{2}
    + \frac{3}{8}\bigl(4x\cos\delta + 4x^{2}\bigr)^{2}
    + O(x^{3}) \nonumber\\[3pt]
&= 1 - 2x\cos\delta + 2x^{2}(3\cos^{2}\delta -1) + O(x^{3}) ,
\label{eq:den_expand_inverse}
\end{align}

\noindent and Eq.~(\ref{eq:numerical_beta_min}) becomes

\begin{align}
\cos\beta_{\min}&\approx\big[1 + 2x\cos\gamma + 4x^2(\cos\gamma\cos\delta+\sin\gamma\sin\delta - 1)\big]
\big[1 - 2x\cos\delta + x^2(6\cos^2\delta - 2)\big] \nonumber\\
&=1 + 2x(\cos\gamma - \cos\delta)
+ x^2\!\Big[\,4(\cos(\gamma-\delta)-1) - 4\cos\gamma\cos\delta + (6\cos^2\delta - 2)\Big]
+ O(x^3).
\label{eq:cosbmin_expand_final}
\end{align}

This expression can be simplified and one finds
\begin{equation}
\cos\beta_{\min}\simeq1+ 2x(\cos\gamma - \cos\delta)+ x^2\!\Big[4\cos(\gamma-\delta)- 4\cos\gamma\cos\delta+ 6\cos^2\delta- 6\Big]+ O(x^3).
\label{eq:cosbmin_final}
\end{equation}
If only the leading contribution is retained, Eq.~\eqref{eq:cosbmin_final} reduces to the simple linear form
\begin{align}
\cos\beta_{\min} &\approx 1 - 2x(\cos\delta - \cos\gamma) + O(x^2),\\
&=1-2x[\frac{1}{2\kappa^2}\cos\delta-\frac{1}{\kappa}\sqrt{1-\frac{1}{4\kappa^2}}\sin\delta]
\label{eq:cosbmin_linear}
\end{align}

\noindent
where in the last step we used Eq.~(\ref{eq:cosgamma}).
On thus recognizes that in the limit of small surface roughness $x=R_a/R_c$, the deviation of $\beta_{\min}$ is primarily determined by the difference between $\cos\delta$ and $\cos\gamma$.

The final result for $\beta_{\min}$ is thus

\begin{align}
\boxed{\beta_{\min}\approx\arccos(1-2x[\frac{1}{2\kappa^2}\cos\delta-\frac{1}{\kappa}\sqrt{1-\frac{1}{4\kappa^2}}\sin\delta])}
\label{eq:beta_min_approx}
\end{align}

We now can calculate the average distance $\bar{d}_{\rm OC}$ given by the integral Eq.~(\ref{eq:numerical_original}).\\
To obtain the analytical expression of the distance $OC$, we rewrite $|{\bf OC}|$ in terms of the particle coordinates. As illustrated in Fig.~\ref{fig:B7-1}, the required coordinates are given by (Eq.~(\ref{eq:coord_AC}))

\begin{align}
\mathrm{A}:(R_c+2R_a\cos\delta,2R_a\sin\delta),\qquad \mathrm{C}:(R_c+2R_a\cos\delta+R_c\cos(\alpha+\beta),2R_a\sin\delta+R_c\sin(\alpha+\beta))
\end{align}
with
\begin{equation}
\tan\alpha=\frac{2R_a\sin\delta}{R_c+2R_a\cos\delta} \quad .
\end{equation}

Hence we have 
\begin{align}
    L:=|\mathbf{OA}| = \sqrt{R_c^2 + 4R_cR_a\cos\delta + 4R_a^2}
    \label{eq:OA}
\end{align}

\noindent
and the distance $OC$ is obtained as

\begin{align}
    d_\mathrm{OC}(\beta,\delta)=|\mathbf{OC}| &= \sqrt{|OA|^2+|AC|^2+2|OA||AC|\cos\beta}\\
    &=\sqrt{L^2+R_c^2+2LR_c\cos\beta}\\
    &=\sqrt{(L+R_c)^2-4LR_c\sin^2(\beta/2)}\\
    &=(L+R_c)\sqrt{1-m^2(\delta)\sin^2(\beta/2)} 
    \label{eq:d_oc}
\end{align}
where $m^2$ is defined as
\begin{equation}
m^2(\delta)=\frac{4LR_c}{(L+R_c)^2} \quad.
\label{eq:def_m}    
\end{equation}
So the inner integral of Eq.~(\ref{eq:numerical_original}) for the calculation of the average distance between to MPs can be written as:
\begin{align}
\bar{d}_\beta:=    \frac{1}{\beta_{\max} -\beta_{\min}}
\int_{\beta_{\min}}^{\beta_{\max}} d_\mathrm{OC} d\beta &=\frac{(L+R_c)}{\beta_{\max} -\beta_{\min}}
\int_{\beta_{\min}}^{\beta_{\max}}\sqrt{1-m^2\sin^2(\beta/2)}d\beta
\label{eq:inner_line1}\\
    &=\frac{2(L+R_c)}{\beta_{\max} -\beta_{\min}}[E(\beta_{\max}/2|m^2)-E(\beta_{\min}/2|m^2)] \quad .
    \label{eq:inner}
\end{align}

Here $E(.|.)$ is the incomplete elliptic integral of the second kind, which in general cannot be reduced to elementary functions and is therefore difficult to treat analytically.

To carry out this integral we therefore consider the case that $x=R_a/R_c\ll 1$ which allows to advance with this integration. From Eq.~(\ref{eq:OA}) we have
\begin{align}
    L &= R_c\sqrt{1+4x\cos\delta+4x^2}
    \label{eq:L_numerical}\\
    & \approx R_c(1+2x\cos\delta)+O(R_cx^2)
    \label{eq:L_approx}
\end{align}

Inserting this approximation in Eq.~(\ref{eq:def_m}) gives

\begin{align}
    m^2 &= \frac{4L/R_c}{(1+L/R_c)^2}
    \label{eq:m^2_numerical}\\
    & \approx 1-x^2\cos^2\delta+O(x^3) \quad .
    \label{eq:m^2_approx}
\end{align}

So we find that if $x \ll 1$, the term $m^2$ occurring in the integral (\ref{eq:inner_line1}) deviates from unity only by $O(x^2)$, and therefore we can approximate it by 1.0.

Eq.~(\ref{eq:inner})  can thus be written as:

\begin{align}
    \bar{d}_\beta &=\frac{(L+R_c)}{\beta_{\max}-\beta_{\min}}
    \int_{\beta_{\min}}^{\beta_{\max}}\sqrt{1-\sin^2(\beta/2)}\;d\beta \quad,
    \label{eq:inner_2}
\end{align}

which gives, using Eq.~(\ref{eq:L_approx},
\begin{align}
    \bar{d}_\beta= \frac{4R_c(1+x\cos\delta)}{\beta_{\max}-\beta_{\min}}[\sin(\beta_{\max}/2)-\sin(\beta_{\min}/2)]+O(R_cx^2) \quad .
    \label{eq:inner_3}
\end{align}

The integration limits are, Eqs.~(\ref{eq_beta_max_approx}) and \ref{eq:beta_min_approx})
\begin{align}
\beta_{\max}
&=2x\!\left(\frac{1}{\kappa}-\sin\delta\right),
\label{eq:beta_max}
\\[4pt]
\beta_{\min}&=\arccos\!\left(1-2xM(\delta)\right),
\label{eq:beta_min}
\end{align}
where 
\begin{align}
\boxed{ M(\delta)
=\frac{1}{2\kappa^{2}}\cos\delta-\frac{1}{\kappa}
\sqrt{1-\frac{1}{4\kappa^{2}}}\,\sin\delta} .
\label{eq:A_def}
\end{align}

For small $x$ we expand
\begin{align}
 {\sin(\beta_{\max}/2) = \sin\!\left(x\left(\tfrac{1}{\kappa}-\sin\delta\right)\right)}
\label{eq:sin_beta_max_numerical}
\end{align}
and obtain
\begin{align}
 {\sin(\beta_{\max}/2) =x\left(\tfrac{1}{\kappa}-\sin\delta\right)+O(x^{3}),}
\label{eq:sin_beta_max_approx}
\end{align}
and using the half–angle identity
\begin{align}
\sin\!\left(\frac12\arccos(1-2xM)\right)&=\sqrt{xM(\delta)} \quad ,
\label{eq:half_angle_simplify}
\end{align}

Results that the second factor in Eq.~(\ref{eq:inner_3}) becomes
\begin{align}
\sin\!\left(\beta_{\max}/2\right)-\sin\!\left(\beta_{\min}/2\right)&=x\left(\tfrac{1}{\kappa}-\sin\delta\right)-\sqrt{xM(\delta)}+O(x^{3}).
\label{eq:num_expand}
\end{align}

Next we expand the denominator:
\begin{align}
\beta_{\max}-\beta_{\min}
&=
2x\!\left(\tfrac{1}{\kappa}-\sin\delta\right)
-
\arccos\!\left(1-2xM(\delta)\right).
\label{eq:den_raw}
\end{align}

To expand the $\arccos$ term we use the small–argument expansion
\begin{align}
\arccos(1-\varepsilon)=\sqrt{2\varepsilon}
+\frac{(2\varepsilon)^{3/2}}{24}
+O(\varepsilon^{5/2}), 
\qquad \varepsilon\to 0,
\label{eq:arccos_expand}
\end{align}
and with $\varepsilon=2xM(\delta)$ this becomes
\begin{align}
\arccos(1-2xM)&=2\sqrt{xM(\delta)}+\frac{(4xM(\delta))^{3/2}}{24}+O(x^{5/2}).
\label{eq:arccos_result}
\end{align}

Thus the denominator takes the form
\begin{align}
\beta_{\max}-\beta_{\min}&=2x\!\left(\tfrac{1}{\kappa}-\sin\delta\right)-2\sqrt{xM(\delta)}-\frac{(4xM(\delta))^{3/2}}{24}+O(x^{2}),
\\[4pt]
&=2\!\left[x\!\left(\tfrac{1}{\kappa}-\sin\delta\right)-\sqrt{xM(\delta)}\right]-\frac{1}{3}(xM(\delta))^{3/2}+O(x^{2}),
\label{eq:den_final}
\end{align}
where we have collected all terms of order $x^{3/2}$.

Next we consider in Eq.~(\ref{eq:inner_3}) the ratio
\begin{align}
\frac{
\sin(\beta_{\max}/2)-\sin(\beta_{\min}/2)}{\beta_{\max}-\beta_{\min}}.
\end{align}
Using the expansions derived above, i.e., Eqs.~(\ref{eq:num_expand}) and (\ref{eq:den_final}), we obtain 

\begin{align}
\frac{\sin(\tfrac{\beta_{\max}}{2})-\sin(\tfrac{\beta_{\min}}{2})}{\beta_{\max}-\beta_{\min}}&=\frac{x(\frac{1}{\kappa}-\sin\delta)-\sqrt{xM(\delta)}}{2[x(\frac{1}{\kappa}-\sin\delta)-\sqrt{xM(\delta)}]-\frac{1}{3}(xM(\delta))^{3/2}}\\
&=\frac{1}{2}+\frac{\frac{1}{3}(xM(\delta))^{3/2}}{2[x(\frac{1}{\kappa}-\sin\delta)-\sqrt{xM(\delta)}]-\frac{1}{3}(xM(\delta))^{3/2}}
\label{eq:ratio_final_right}
\end{align}

Substituting this expression into Eq.~(\ref{eq:inner_3}) gives\\
\begin{align}
\bar{d}_\beta&=2R_c(1+x\cos\delta)+\frac{2R_c(1+x\cos\delta)(Mx)^{3/2}}{3[x(\frac{1}{\kappa}-\sin\delta)-\sqrt{xM(\delta)}]-\frac{1}{2}(xM(\delta))^{3/2}}\\
&=2R_c(1+x\cos\delta)-\frac{2R_c(1+x\cos\delta)Mx}{3}*\frac{1}{1-\frac{(\frac{1}{\kappa}-\sin\delta)}{\sqrt{M}}\sqrt{x}+\frac{1}{6}Mx}
\label{eq:inner_avg_noparam}
\end{align}
Using the relation
\begin{equation}
    \frac{1}{1+au+bu^2} = 1-au+(a^2-b)u^2+(-a^3+2ab)u^3+O(x^4)
\end{equation}

\noindent
for the case $u=\sqrt{x}$ one finds
    
\begin{align}
    \bar{d}_\beta&=2R_c(1+x\cos\delta)-\frac{2R_c(1+x\cos\delta)Mx}{3}\left(1+\frac{\frac{1}{\kappa}-\sin\delta}{\sqrt{M}}\sqrt{x}+(\frac{(\frac{1}{\kappa}-\sin\delta)^2}{M}-\frac{M}{6})x\right)
\end{align}

thus

\begin{align}
\boxed{\bar{d}_\beta=2R_c+2R_cx(\cos\delta-\frac{M}{6})-2R_cx^{3/2}\frac{(\frac{1}{\kappa}-\sin\delta)\sqrt{M}}{3}+O(x^2)}
\label{eq:inner_avg_noparam}
\end{align}

The average distance between two MPs is given by, see Eq.~(\ref{eq:numerical_original}), 
\begin{align}
\bar{d}=\frac{1}{2\delta_{\max}}\int_{-\delta_{\max}}^{\delta_{\max}}\bar{d}_\beta\, d\delta 
\end{align}

\begin{align}
\bar{d}&=\frac{1}{2\delta_{\max}}\int_{-\delta_{max}}^{\delta_{\max}}
\left[2R_c+2R_cx(\cos\delta-\frac{M}{6})-2R_cx^{3/2}\frac{(\frac{1}{\kappa}-\sin\delta)\sqrt{M}}{3}\right] d\delta+O(R_c x^{2}).
\label{eq:outer_noparam_split}
\end{align}

with

\begin{equation}
    M(\delta)
=\frac{1}{2\kappa^{2}}\cos\delta-\frac{1}{\kappa}
\sqrt{1-\frac{1}{4\kappa^{2}}}\,\sin\delta
\end{equation}

Using, see Eq.~(\ref{eq_delta_max_approx},
\begin{align}
\cos\delta_{\max}=\sqrt{1-\tfrac{1}{4\kappa^{2}}},
\qquad
\sin\delta_{\max}=\frac{1}{2\kappa},
\end{align}
we get the result
\begin{align}
\bar{d} &= 2R_c + \frac{R_cx \sin\delta_{\max}}{\delta_{\max}} \left( 2 - \frac{1}{6\kappa^2} \right) - \frac{R_cx^{3/2}}{3\kappa \delta_{\max}} J_1 + \frac{R_cx^{3/2}}{3\delta_{\max}} J_2 + O(x^2) \\
J_1 &= \int_{-\delta_{\max}}^{\delta_{\max}} \sqrt{ \frac{1}{\kappa} \cos(\delta - \phi) } \, d\delta \\
J_2 &= \int_{-\delta_{\max}}^{\delta_{\max}} \sin\delta \, \sqrt{ \frac{1}{\kappa} \cos(\delta - \phi) } \, d\delta \\
\phi &= \arccos\left( \frac{1}{2\kappa} \right), \quad \delta_{\max} = \frac{\pi}{2} - \phi
\end{align}

If we simplify the formula we obtain
\begin{equation}
    \bar{d} = 2R_c\left[1+(1-\frac{1}{12\kappa^2})\frac{\sin\delta_{max}}{\delta_{max}}x+\frac{(\kappa J_2-J_1)}{6\kappa\delta_{max}}x^{\frac{3}{2}}\right]+O(x^2) \quad .
    \label{eq:d_expression}
\end{equation}

As mentioned in the main text, we introduce an effective radius
\begin{equation}
R_{\rm eff}=R_c\left(1+B\,\frac{R_a}{R_c}\right).
\end{equation}
In the present calculation we identify the effective diameter with the
average center--to--center distance at contact, i.e.,
\begin{equation}
\bar d \equiv 2R_{\rm eff}
=2R_c\left(1+B\,\frac{R_a}{R_c}\right)
=2R_c+2B\,R_a.
\end{equation}
Hence, by comparing the expansion of $\bar d$ with the above expression, the factor $B$ can be extracted: 

\begin{align}
    B &= \frac{d(\bar{d})}{dx} = \frac{ \sin\delta_{max}}{2\delta_{max}}\left(2-\frac{1}{6\kappa^2}\right)-\frac{J_1}{4\kappa\delta_{max}}x^{1/2}+\frac{J_2}{4\delta_{max}}x^{1/2}\\
    &=\frac{1}{4\kappa\arcsin(\frac{1}{2\kappa})}\left(2-\frac{1}{6\kappa^2}\right)-\frac{J_1}{4\kappa\delta_{max}}x^{1/2}+\frac{J_2}{4\delta_{max}}x^{1/2}
    \label{eq_B_exact}
\end{align}

If $x \ll 1$, this expression simplifies to
\begin{equation}
    B \approx \frac{1}{4\kappa\arcsin(\frac{1}{2\kappa})}\left(2-\frac{1}{6\kappa^2}\right) \quad.
    \label{eq_B_approx}
\end{equation}

This relation corresponds to Eq.~(4) in the main text.

\begin{figure}[ht]
\centering
\includegraphics[width=13cm]{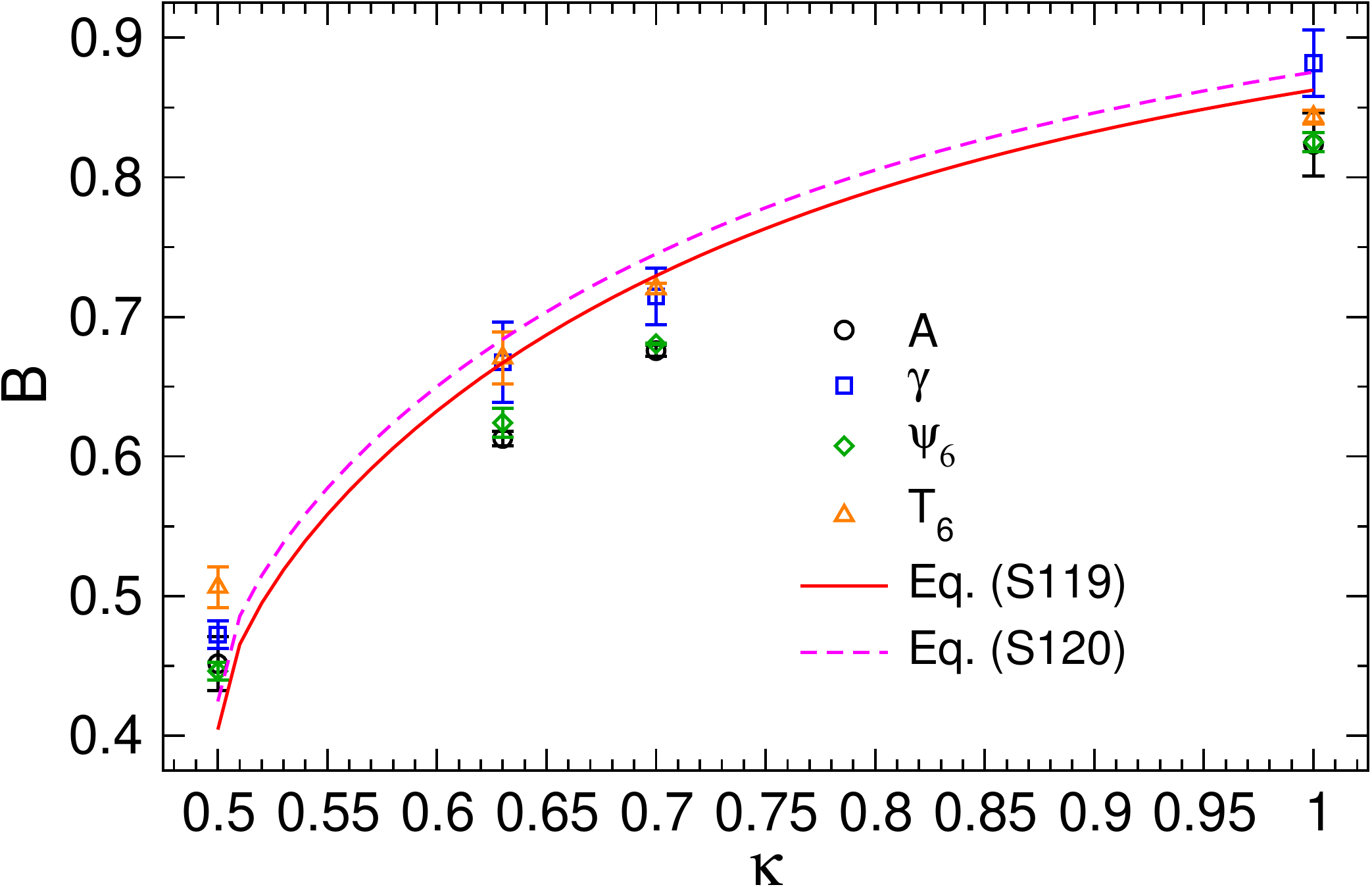}
\caption{Scaling factor $B$ as a function of the coverage ratio $\kappa$. Symbols show $B(\kappa)$ obtained independently from the collapse of the structural observables $A$, $\psi_6$, $T_6$, and the Lindemann ratio $\gamma$. The solid and dashed lines correspond to the theoretical predictions from Eq.~\ref{eq_B_exact} and Eq.~\ref{eq_B_approx}, respectively. }
    \label{fig_B_kappa_eq125}
\end{figure}

Equation~\ref{eq_B_exact} follows from differentiating $\bar d$ with respect to $x$ and thus contains the subleading corrections $\propto x^{1/2}$ originating from the $O(x^{3/2})$ terms in $\bar d$ (through the integrals $J_1$ and $J_2$). In contrast, Eq.~(\ref{eq_B_approx}) corresponds to the leading-order, linear-in-$x$ contribution and is obtained by taking the limit $x\to 0$, i.e. by neglecting the $x^{1/2}$ terms in Eq.~(\ref{eq_B_exact}). To evaluate Eq.~(\ref{eq_B_exact}) 
numerically we use a representative small value $x=5 \cdot 10^{-3}$, which accurately approximates the $x\to 0$ limit while keeping the expressions well defined; further decreasing $x$ does not change the resulting curve within numerical accuracy. We use this $x\to 0$ estimate as the geometrical prediction for the rescaling factor shown in Fig.~\ref{fig_B_kappa_eq125} and Fig.~4 in the main text. 

\begin{figure}[ht]
\centering
\includegraphics[width=13cm]{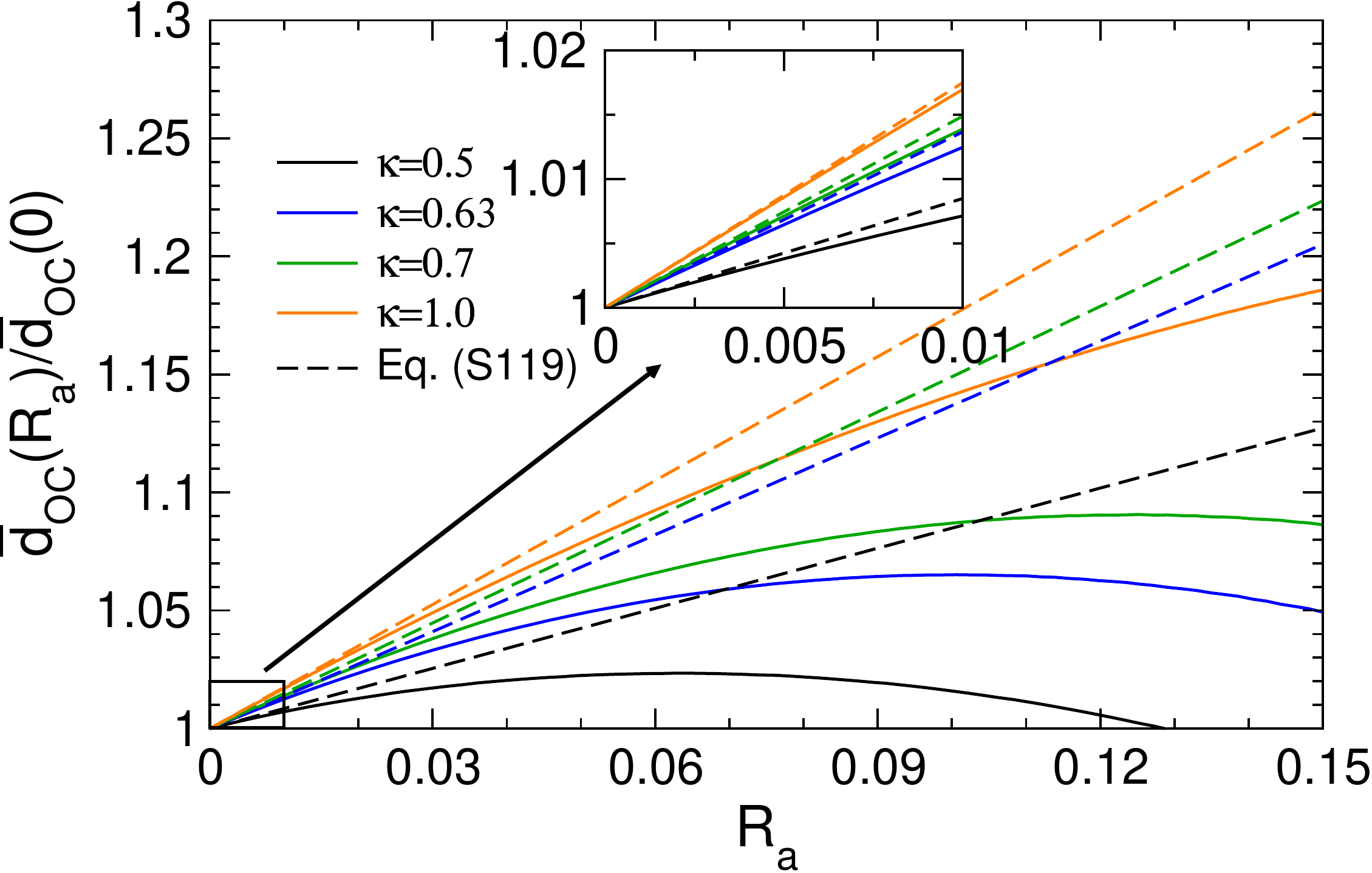}
\caption{Normalized mean contact distance $\bar{d}_{OC}(R_a,\kappa)/\bar{d}_{OC}(0,\kappa)$ as a function of the asperity radius $R_a$ for different coverages $\kappa$. The solid lines show the exact numerical evaluation of Eq.~\eqref{eq:numerical_original}, while the dashed lines show the analytical approximation from Eq.~\eqref{eq_B_exact}, evaluated at $x=0.005$. The inset enlarges the small-$R_a$ regime. This figure is intended to compare the exact numerical result of Eq.~\eqref{eq:numerical_original} with its analytical approximation. One sees that the agreement is good in the small-roughness regime, whereas visible deviations appear at larger $R_a$, in particular for small $\kappa$.
}

    \label{fig:d_Ra_kappa}
\end{figure}

Figure~\ref{fig:d_Ra_kappa} shows the mean contact distance $\bar{d}_{OC}(R_a,\kappa)$ between two touching MPs, obtained from the exact numerical evaluation of Eq.~\ref{eq:numerical_original}. This exact numerical result can be compared with the analytical small-roughness calculation discussed above Eq.~\ref{eq_B_exact}. 
As can be seen from this figure, the exact numerical result for $\bar{d}_{OC}(R_a,\kappa)$ is not perfectly linear over the full range of $R_a$; in particular, for larger asperities the increase becomes less linear, and for smaller $\kappa$ the curves may even show a tendency towards saturation. However, in the small-roughness regime the dependence is reasonably well approximated by a linear form, confirming the accuracy of Eq.~(\ref{eq_B_exact}) in that limit. 

\end{widetext}
\end{document}